\documentclass[twocolumn]{aastex61}
\pdfoutput=1 
\usepackage{amsmath,amstext,mathtools}
\usepackage[T1]{fontenc}
\usepackage{apjfonts} 
\usepackage[figure,figure*]{hypcap}
\usepackage{changepage}
\usepackage{multirow}
\usepackage{rotating}
\graphicspath{{plots/}}

\shorttitle{Gamma-Rays from Kilonova}
\shortauthors{Korobkin et al.}

\newcommand{\bmd}{$\beta^{-}$-decay}

\newcommand{\gray}{$\gamma$-ray}
\newcommand{\grays}{$\gamma$-rays}

\newcommand{\Am}[1]{$\prescript{#1}{95}{\text{Am}}$}

\newcommand{\Bi}[1]{$\prescript{#1}{83}{\text{Bi}}$}
\newcommand{\Bk}[1]{$\prescript{#1}{97}{\text{Bk}}$}
\newcommand{\Ce}[1]{$\prescript{#1}{58}{\text{Ce}}$}

\newcommand{\Cm}[1]{$\prescript{#1}{96}{\text{Cm}}$}
\newcommand{\Cs}[1]{$\prescript{#1}{55}{\text{Cs}}$}
\newcommand{\Eu}[1]{$\prescript{#1}{63}{\text{Eu}}$}
\newcommand{\I}[1]{$\prescript{#1}{53}{\text{I}}$}

\newcommand{\La}[1]{$\prescript{#1}{57}{\text{La}}$}
\newcommand{\Nb}[1]{$\prescript{#1}{41}{\text{Nb}}$}
\newcommand{\Np}[1]{$\prescript{#1}{93}{\text{Np}}$}
\newcommand{\Os}[1]{$\prescript{#1}{76}{\text{Os}}$}
\newcommand{\Pa}[1]{$\prescript{#1}{91}{\text{Pa}}$}
\newcommand{\Pb}[1]{$\prescript{#1}{82}{\text{Pb}}$}

\newcommand{\Pt}[1]{$\prescript{#1}{84}{\text{Pt}}$}

\newcommand{\Ra}[1]{$\prescript{#1}{88}{\text{Ra}}$}

\newcommand{\Sb}[1]{$\prescript{#1}{51}{\text{Sb}}$}
\newcommand{\Sn}[1]{$\prescript{#1}{50}{\text{Sn}}$}
\newcommand{\Sr}[1]{$\prescript{#1}{38}{\text{Sr}}$}

\newcommand{\Th}[1]{$\prescript{#1}{90}{\text{Th}}$}
\newcommand{\U}[1]{$\prescript{#1}{92}{\text{U}}$}
\newcommand{\Xe}[1]{$\prescript{#1}{54}{\text{Xe}}$}
\newcommand{\Y}[1]{$\prescript{#1}{39}{\text{Y}}$}
\newcommand{\Zr}[1]{$\prescript{#1}{40}{\text{Zr}}$}

\begin{document}

\title{Gamma-Rays from Kilonova: A Potential Probe of $r$-process Nucleosynthesis}

\author[0000-0003-4156-5342]{Oleg Korobkin}
\affiliation{Center for Theoretical Astrophysics, Los Alamos National Laboratory, Los Alamos, NM 87545, USA}
\affiliation{Joint Institute for Nuclear Astrophysics - Center for the Evolution of the Elements, USA}
\affiliation{Computer, Computational, and Statistical Sciences Division, Los Alamos National Laboratory, Los Alamos, NM 87545, USA}

\author[0000-0001-6893-0608]{Aimee M.~Hungerford}
\affiliation{Center for Theoretical Astrophysics, Los Alamos National Laboratory, Los Alamos, NM 87545, USA}
\affiliation{Joint Institute for Nuclear Astrophysics - Center for the Evolution of the Elements, USA}
\affiliation{X Computational Physics Division, Los Alamos National Laboratory, Los Alamos, NM 87545, USA}

\author[0000-0003-2624-0056]{Christopher~L. Fryer}
\affiliation{Center for Theoretical Astrophysics, Los Alamos National Laboratory, Los Alamos, NM 87545, USA}
\affiliation{Joint Institute for Nuclear Astrophysics - Center for the Evolution of the Elements, USA}
\affiliation{Computer, Computational, and Statistical Sciences Division, Los Alamos National Laboratory, Los Alamos, NM 87545, USA}
\affiliation{The University of Arizona, Tucson, AZ 85721, USA}
\affiliation{Department of Physics and Astronomy, The University of New Mexico, Albuquerque, NM 87131, USA}
\affiliation{The George Washington University, Washington, DC 20052, USA}

\author[0000-0002-9950-9688]{Matthew~R. Mumpower}
\affiliation{Center for Theoretical Astrophysics, Los Alamos National Laboratory, Los Alamos, NM 87545, USA}
\affiliation{Joint Institute for Nuclear Astrophysics - Center for the Evolution of the Elements, USA}
\affiliation{Theoretical Division, Los Alamos National Laboratory, Los Alamos, NM 87545, USA}

\author[0000-0002-9950-9688]{G.~Wendell Misch}
\affiliation{Center for Theoretical Astrophysics, Los Alamos National Laboratory, Los Alamos, NM 87545, USA}
\affiliation{Joint Institute for Nuclear Astrophysics - Center for the Evolution of the Elements, USA}
\affiliation{Theoretical Division, Los Alamos National Laboratory, Los Alamos, NM 87545, USA}

\author[0000-0002-4375-4369]{Trevor~M. Sprouse}
\affiliation{University of Notre Dame, Notre Dame, IN 46556, USA}

\author[0000-0002-5936-3485]{Jonas Lippuner}
\affiliation{Center for Theoretical Astrophysics, Los Alamos National Laboratory, Los Alamos, NM 87545, USA}
\affiliation{Joint Institute for Nuclear Astrophysics - Center for the Evolution of the Elements, USA}
\affiliation{Computer, Computational, and Statistical Sciences Division, Los Alamos National Laboratory, Los Alamos, NM 87545, USA}

\author[0000-0002-4729-8823]{Rebecca Surman}
\affiliation{Joint Institute for Nuclear Astrophysics - Center for the Evolution of the Elements, USA}
\affiliation{University of Notre Dame, Notre Dame, IN 46556, USA}

\author[0000-0002-0861-3616]{Aaron~J. Couture}
\affiliation{Joint Institute for Nuclear Astrophysics - Center for the Evolution of the Elements, USA}
\affiliation{Physics Division, Los Alamos National Laboratory, Los Alamos, NM 87545, USA}

\author[0000-0002-6664-4306]{Peter F. Bloser}
\affiliation{Intelligence and Space Research Division, Los Alamos National Laboratory, Los Alamos, NM 87545, USA}

\author[0000-0000-0000-0000]{Farzane Shirazi}
\affiliation{Department of Physics, University of New Hampshire, Durham, NH 03824, USA}

\author[0000-0002-5412-3618]{Wesley~P. Even}
\affiliation{Center for Theoretical Astrophysics, Los Alamos National Laboratory, Los Alamos, NM 87545, USA}
\affiliation{Joint Institute for Nuclear Astrophysics - Center for the Evolution of the Elements, USA}
\affiliation{Computer, Computational, and Statistical Sciences Division, Los Alamos National Laboratory, Los Alamos, NM 87545, USA}
\affiliation{Department of Physical Science, Southern Utah University, Cedar City, UT 84720, USA}

\author[0000-0001-7120-7234]{W. Thomas Vestrand}
\affiliation{Intelligence and Space Research Division, Los Alamos National Laboratory, Los Alamos, NM 87545, USA}

\author[0000-0003-4496-7128]{Richard S. Miller}
\affiliation{Johns Hopkins University, Applied Physics Laboratory, Laurel, MD 20723, USA}

\begin{abstract}
The mergers of compact binaries with at least one neutron star component are the potential leading sites of the production and ejection of $r$-process elements. Discoveries of galactic binary pulsars, short gamma-ray bursts, and gravitational-wave detections have all been constraining the rate of these events, while the gravitational wave plus broadband electromagnetic coverage of binary neutron star merger (GW170817) has also placed constraints on the properties (mass and composition) of the merger ejecta. But uncertainties and ambiguities in modeling the optical and infrared emission make it difficult to definitively measure the distribution of heavy isotopes in these mergers. In contrast, gamma rays emitted in the decay of these neutron-rich ejecta may provide a more direct measurement of the yields. We calculate the gamma production in remnants of neutron star mergers, considering two epochs: a kilonova epoch, lasting about two weeks, and a much later epoch of tens and hundreds of thousands of years after the merger. For the kilonova epoch, when the expanding ejecta is still only partially transparent to gamma radiation, we use 3D radiative transport simulations to produce the spectra. We show that the gamma-ray spectra associated with beta- and alpha-decay provide a fingerprint of the ejecta properties and, for a sufficiently nearby remnant, may be detectable, even for old remnants. We compare our gamma spectra with the potential detection limits of next generation detectors, including the \emph{Lunar Occultation Explorer (LOX)}, the \emph{All-sky Medium Energy Gamma-ray Observatory (AMEGO)}, and the \emph{Compton Spectrometer and Imager (COSI)}. We show that fission models can be discriminated via the presence of short-lived fission fragments in the remnant spectra.
\end{abstract}


\section{Introduction}

Over four decades ago, \cite{lattimer74} proposed that rapid decompression of neutron-rich matter from a tidally disrupted neutron star could account for the $r$-process production of the universe. Proving this point requires demonstrating that the rate of neutron star mergers is sufficiently high and that the cumulative nucleosynthetic yield is plentiful, given the merger rate, and furthermore, produces the solar-like distribution in proper agreement with $r$-process enriched metal-poor stars~\citep{sneden96,beers05,hansen18,ji18}. Rates of these mergers from theoretical~\citep[e.g.][]{fryer99a, dominik12} and observed binary pulsars~\citep[e.g.][]{kalogera04, chen13}, and gamma-ray bursts~\citep[GRBs; e.g.][]{paul18} span a wide range, arguing that they produce between $<1$\% and 100\% of the $r$-process~\citep{cote17}.  Theoretical rates are uncertain because binary population synthesis models suffer from large uncertainties in stellar evolution (e.g. stellar radii and shell sizes), binary evolution (e.g. common envelope evolution and mass transfer) and supernova (e.g. neutron star kicks) properties.  Observations, on the other hand, are prone to bias (e.g. determining the completeness of the observed sample).  The gravitational-wave detection of GW170817 provided an independent observational constraint, arguing for a sufficiently high rate that, with yields currently predicted by simulations, mergers could produce most, if not all, of the $r$-process elements~\citep{cote18, rosswog18}. While ongoing gravitational-wave detections are refining these rate estimates, studies from the perspective of galactic chemical evolution indicate that several $r$-process sites were operating in the early universe~\citep{hotokezaka18,cote19,simonetti19}.

With the merger rate increasingly constrained, the viability of mergers as an $r$-process source depends more upon the uncertainties in the amount and composition of the merger ejecta. The ejecta from the merger occurs during the initial tidal disruption, as well as at late times, as the debris accretes onto the merged core~\citep{dessart09a, perego14a, martin15, siegel17}. Theory predicts a range of ejecta masses about ${10^{-3} - 10^{-2}M_\odot}$~\citep{korobkin12, bauswein13, hotokezaka13, endrizzi16, radice16, sekiguchi16}. While the tidal (or dynamical) ejecta is believed to be neutron rich, and hence has been argued to produce a composition that is "robust" in $r$-process elements, the neutron fraction can be reset by neutrinos, producing everything from iron peak elements to the heavy $r$-process~\citep{fernandez13, wanajo14, fernandez15, just15}. To truly understand the yields from neutron star mergers, we must understand both the ejecta composition and their amount.

Optical, ultraviolet and infrared electromagnetic counterparts of neutron star mergers provide one venue for inferring the nature of these ejecta~\citep{li98,piran05,metzger10a}. Specifically, astronomers argued for both ``red'' (produced from ejecta with heavy $r$-process) and ``blue'' (ejecta with atomic masses only up to and including the second $r$-process peak) components~\citep{metzger12}. The Lanthanides synthesized as part of the heavy $r$-process have many lines in the ultraviolet, optical, and near-infrared wavelength bands, driving the emission to the mid-infrared.  These ejecta produce the "red" component in the emission seen in many calculations~\citep[e.g.][]{kasen13,tanaka13,fontes15}. If the late-time ejecta is less neutron rich to the point that there are insufficient neutrons to produce the heavy $r$-process elements, the ejecta will generate a bright, short-lived blue transient \citep[e.g.][]{metzger12,barnes13,perego14a,wollaeger18}.  

Prior to GW170817, astronomers had to make a series of assumptions to probe the ejecta properties of neutron star mergers.   First, they established a connection between short GRBs and neutron star mergers by observing that offset distributions~\citep{fong13} of short GRBs match predictions of neutron star populations~\citep{bloom99,fryer99a}. They then assumed that deviations in the power-law decay of GRB afterglows could arise from the emergence of radioactive emission from the ejecta. A number of kilonova candidates were identified~\citep{perley09,tanvir13,fong14,jin15,jin16,lamb19}. However, observing such components is difficult, because the kilonova light-curve signal must be separated from much brighter background of the GRB afterglow, and shocks in the afterglow may produce bumps in the optical/infrared that can be mistaken for kilonova light~\citep{kasliwal17}.  If the infrared excess has a corresponding X-ray flare, it is more likely to be caused by shock interactions with the inhomogeneities in the circumstellar medium rather than powered by the ejecta radioactivity.  With GW170817, the ejecta emission --kilonova-- was observed unambiguously for the first time, providing a first direct probe of this phenomenon.  The combined strong blue and red components of this merger seemed to fit the models predicted for both dynamical/tidal and late-time wind/disk ejecta, allowing to infer the masses of individual components.

But recent analysis of the GW170817 kilonova spectra has made it clear that uncertainties in the model would make it difficult to make concrete claims about the amount and composition of the ejecta. Overviews of the analyses from different groups show a broad range of inferred ejecta masses~\citep{cote18,ji19}.  Much of the uncertainty in light-curve calculation comes from the modeling of opacities and their incorporation into transport codes~\citep{kasen13,tanaka13,fontes17}. The methods used to calculate the opacities, the number of levels (and lines) considered, and the methods to combine these opacities in an expanding medium all can affect the light curve~\citep{fontes19}. However, the uncertainties in the ejecta properties (density, velocity, and composition distributions) and morphology produce even larger uncertainties~\citep{grossman14,wollaeger18}.  Thus, even with the pristine data from GW170817, it is difficult to determine the ejecta masses to better than an order of magnitude.  Other effects also muddle the interpretation and analysis of the kilonova emission.  For example, the flux (especially in the optical and ultraviolet) can vary dramatically with the viewing angle~\citep[see, e.g.][]{fernandez17, wollaeger18}.  All of these studies assumed that radioactive decay powers the emission, but additional energy sources (a pulsar or emission from accretion onto the compact remnant) can also impact the light curve~\citep{wollaeger19}. 

With all of these uncertainties, it is difficult to estimate accurate ejecta masses based solely on the broadband light curves. Obtaining detailed abundances is even more challenging. It is possible that spectral features can provide evidence of the composition, and there are hints that the GW170817 must have ejected at least some light $r$-process elements~\citep{pian17,watson19}, but obtaining detailed yields requires full, time-dependent and out-of-equilibrium opacities. It is also possible to constrain ejecta masses via radioactive heating \citep{piran14,rosswog18}, but this approach is only partially successful, as heating is expected to look similar for many initial conditions~\citep{lippuner15, eichler19}.

As with $^{56}$Ni yields in thermonuclear supernovae~\citep{churazov14,the14} and $^{56}$Ni and $^{44}$Ti yields in core-collapse supernovae \citep{hungerford05, grefenstette14, grefenstette17}, a more direct measurement of the yields can be obtained by observing the photons from the decay of radioactive nuclei in the ejecta. In this paper, we study the potential of measurements of decay photons to probe the nucleosynthetic yields and nuclear physics in neutron star mergers.  We focus our efforts on the study of \grays{} produced by the nuclear decay of neutron-rich nuclei. In a pioneering work on this subject, \cite{hotokezaka16} calculated the \gray{} signal from kilonova ejecta and found that it would be detectable out to $\sim 3-10\, {\rm Mpc}$ with current detectors. However, their work was done without modeling \gray{} transport, which can significantly redistribute emission to lower energies, impairing detectability. Recently, \cite{li19} constructed a semianalytic model of the radioactive \gray{} emission from kilonovae powered by nuclear decays. An earlier study by~\cite{janiuk14} suggested detection of X-ray emission from iron-group isotopes synthesized in central engines of GRBs.

Although the main peak flux of \grays{} happens at early times, the emission continues for more than a hundred thousand years after the merger. Therefore, it is possible that there is a nearby kilonova remnant that can be observed. In a complementary study, \cite{wu19} consider prospects of finding such neutron star merger remnants in the Milky Way galaxy. In earlier work, \cite{qian98} concluded that sensitive \gray{} detector can observe lines from a few long-lived heavy radioactive isotopes decaying in supernova remnants, in particular \Sb{125}, \Sn{126}, \Cs{137}, \Ce{144}, \Eu{155} and \Os{194}. Subsequently, \cite{ripley14} investigated search prospects for both supernova and neutron star merger remnants within the Galactic plane using the NuGRID and LOFT X-ray observatories. It was found that $>10^2$ overabundance is required to detect the lines of the most promising isotope, \Sn{126}. \cite{fuller19} argued that thermal positron production at the initial stage of kilonova explosion could generate strong 511~keV annihilation line signature which might help with identifying such remnants. We further explore possible \gray{} emission from the remnants, using detailed $r$-process nucleosynthesis calculations and models for ejecta deceleration in the interstellar medium.
  
Section~\ref{sec:nuclear} introduces our method, including the ejecta morphologies, detailed nucleosynthesis models, and \gray{} source calculation.  Section~\ref{sec:transients} discusses early-time \gray{} signatures of kilonova, following the fully 3D transport of the emitted \grays{} through the ejecta. In section~\ref{sec:remnants}, we calculate the properties of neutron star merger remnants. We conclude with a comparison with upcoming \gray{} missions.

\section{Gamma Rays from $r$-process Yields}
\label{sec:nuclear}

The neutron-rich ejecta from neutron star mergers are expected to produce a wide range of elements from the iron peak to third $r$-process peak and beyond. Gamma-ray signatures would therefore depend not only on the neutron richness, but also on thermodynamic history and the morphology of the ejecta, which affect this history. The ejecta neutron richness ranges from extremely high in the dynamical part produced in the process of tidal disruption of the neutron stars, to the medium-richness outflows from the accretion disk~\citep{janiuk14,siegel17,miller19} up to the much more symmetric ejecta in the outflows from central merger product~\citep{perego14a,martin15}.

\begin{figure}[htp]
\hspace{-1cm}
\begin{tabular}{c}
\includegraphics[width=0.5\textwidth]{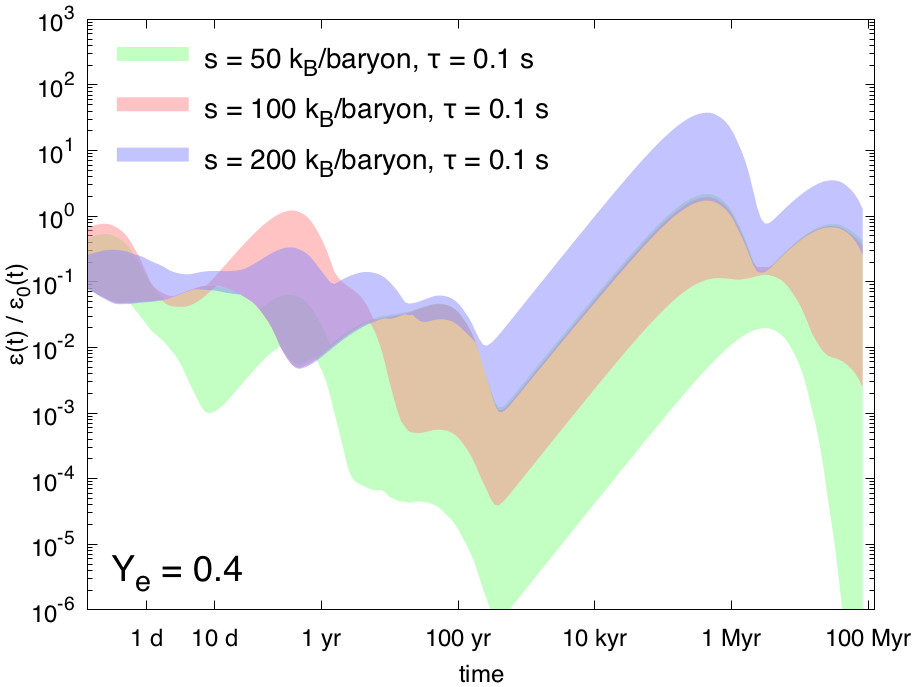} \\
\includegraphics[width=0.5\textwidth]{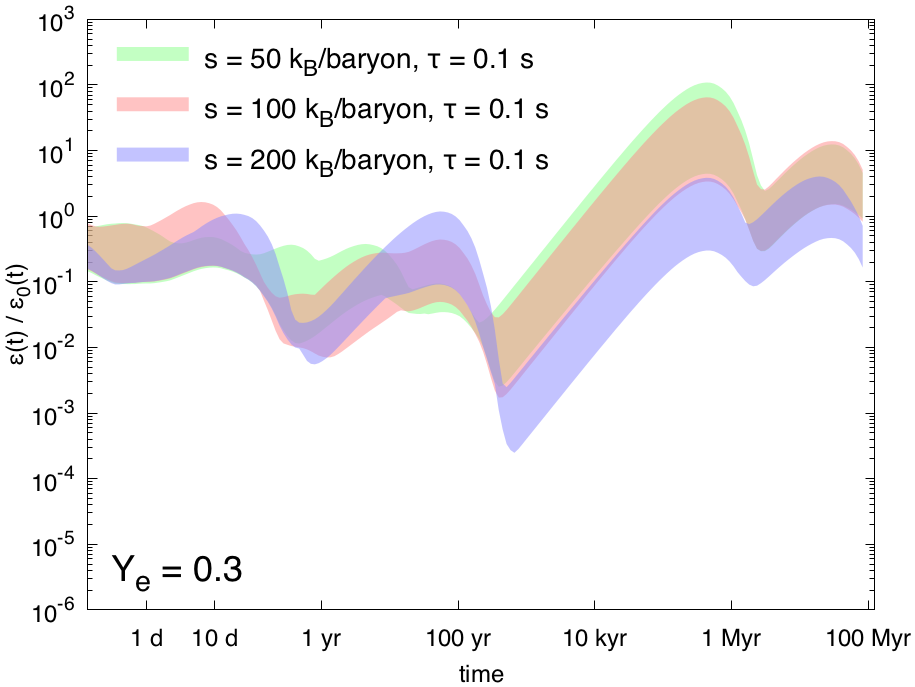} \\
\includegraphics[width=0.5\textwidth]{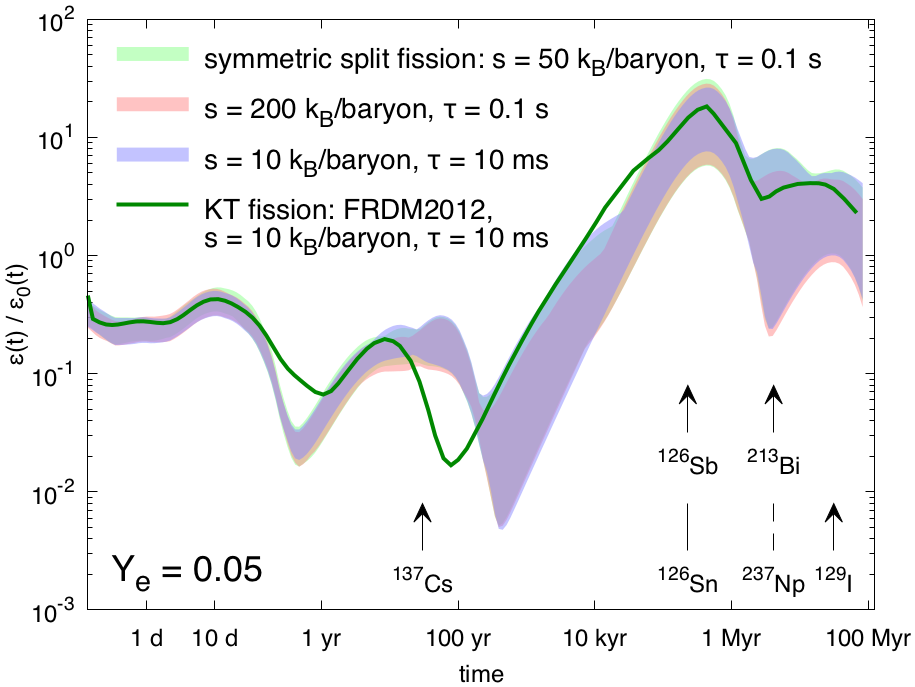}
\end{tabular}
\caption{Normalized rate of nuclear energy produced in $\gamma$-radiation, 
for a range of nuclear mass models. The top, middle, and bottom panels
represent neutron-poor ($Y_e = 0.4$), medium neutron richness ($Y_e=0.3$), 
and neutron-rich ($Y_e = 0.05$) conditions, respectively.
Three colors correspond to different hydrodynamic conditions, encoded in
the expansion timescales $\tau\,[{\rm ms}]$ and starting entropies 
$s\,[k_{\rm B}/{\rm baryon}]$. The rates are normalized to 
${\epsilon_0(t)\sim t^{-1.3}}$. }
\label{fig:gammarates}
\end{figure}

The extent of corresponding variability of the \gray{} production is demonstrated in Figure~\ref{fig:gammarates} which shows it as a function of time for different thermodynamic conditions and nuclear mass models. Here, the $\gamma$-radiation rate $\varepsilon_\gamma(t)$ is normalized to the power-law decay fit~\citep{metzger12,korobkin12} 
    $\varepsilon_0(t) = 2\times10^{10}\,{\rm erg}\,{\rm g}^{-1}{\rm s}^{-1}\;t_d^{-1.3}$,
where the time $t_d$ is measured in days. Such a plot allows forthe easy identification of epochs of active \gray{} emission for a given model.

In Figure~\ref{fig:gammarates}, variability due to nuclear mass model is shown with a color band, while different colors represent different initial hydrodynamic conditions (see details below). The two top panels show composition outcomes from moderately neutron-rich ejecta ($Y_e=0.4$ and $Y_e=0.3$), and exhibit several orders of sensitivity to the hydrodynamics as opposed to only about one order of magnitude sensitivity to nuclear mass model. On the other hand, models starting with extremely neutron-rich conditions ($Y_e=0.05$) exhibit very little variation in \gray{} production (bottom panel). This is because nucleosynthesis in this regime is robust and governed by fission recycling much more than by hydrodynamics~\citep{korobkin12,holmbeck19}.

The power law with fractional power index emerges from a multitude of $\beta^-$-decaying isotopes~\citep{metzger12,hotokezaka17}. At late times, it breaks into individual peaks produced by individual radioactive decay chains. A chain generating peak at time $t$ starts with an isotope having mean lifetime $\tau\approx t$ which does not necessarily produce \grays{}: rather, a different $\beta^-$-decaying isotope downstream with much shorter lifetime may be responsible. Decay chains \Sn{126}$\to$\Sb{126} and \Np{237}$\to\dots$\Bi{213} shown schematically at the bottom panel of Fig.~\ref{fig:gammarates} give examples of such scenario.

To keep parameter space manageable, we only explore a few thermodynamic trajectories representative of different ejecta types and nucleosynthesis models. We adopt a two-component model motivated by numerical simulations of neutron star mergers~\citep{rosswog14} and used in our two-dimensional study of kilonova light curves~\citep{wollaeger18}. As in~\cite{wollaeger18}, the two components are neutron-rich "dynamical ejecta" and lighter $r$-process-producing ``wind''.

\begin{figure}[htp]
\centering
\begin{tabular}{c}
\includegraphics[width=0.45\textwidth]{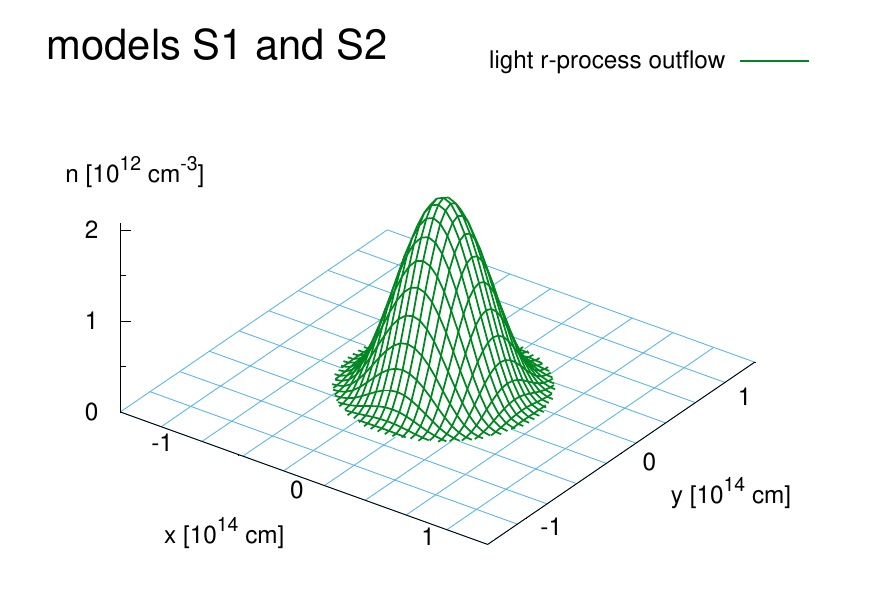} \\
\includegraphics[width=0.45\textwidth]{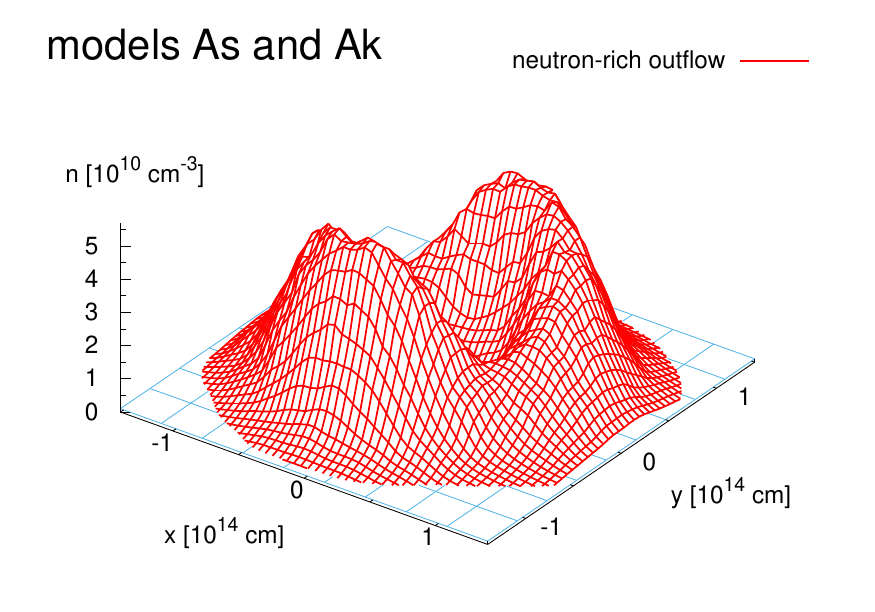}
\end{tabular}
\caption{Ion density at the epoch ${t = 4}$~hr for the two basic 
morphologies used to model early emission: spherical for the wind outflow
(top) and toroidal for the dynamical ejecta (bottom).}
\label{fig:morphologies}
\end{figure}

The morphology of the dynamical ejecta is derived from model A in~\citet[see their Table 1]{rosswog14}, which was computed in the neutron star merger simulation and followed by the subsequent expansion of the ejecta up to homology. We rescale its mass for the best fit to the GW170817 kilonova \citep[as in our models for this event presented in][]{evans17,tanvir17,troja17}. For the secondary, less neutron-rich and wind-like outflow, we pick an analytic spherically symmetric background solution as introduced in \cite{wollaeger18}.  Dynamical ejecta is rescaled to have mass ${m_{\rm dyn}= 0.0065 M_{\odot}}$ and median expansion velocity ${v_{\rm dyn} = 0.2 c}$, while the wind outflow is heavier and slower: ${m_{\rm wind} = 0.03 M_{\odot}}$, ${v_{\rm wind} = 0.08 c}$~\citep{tanvir17}. The morphologies of the two components in our models are depicted in Figure~\ref{fig:morphologies}. 

To produce the nucleosynthetic composition for our model components, we use the Portable Routines for Integrated nucleoSynthesis Modeling (\texttt{PRISM}) reaction network, most recently used in \cite{cote18}, \cite{vassh19}, \cite{sprouse19}. This network uses state-of-the-art nuclear physics inputs \citep[e.g.][]{mumpower17, mumpower18, moller19}, including a consistent treatment of capture rates as well as neutron-induced and $\beta$-delayed fission using the theoretical framework of \cite{kawano08, kawano16}, \cite{mumpower16}. This framework combines together various statistical nuclear model inputs such as nuclear level densities, $\gamma$-ray strength functions and optical potentials to produce well-tested predictions \citep{spyrou16, yokoyama19} for nucleosynthesis calculation. Variations in nuclear binding energies proceed as in~\cite{mumpower15} with the statistical model inputs held fixed.

The time evolution of the abundances $Y_{\rm iso}(t)$ is used to calculate the detailed $\gamma$-ray source. The source represents finely binned spectrum, based on the total spectrum $S(E,t)$, which in turn is computed using abundances of the decaying isotopes, their known $\gamma$-radiation lines and the spectrum of the continuum component (if present), ${\rm RP}_{\rm iso}(E)$:
\begin{align}
    S(E,t) = N_A \sum_{\rm iso} \lambda_{\rm iso} Y_{\rm iso}(t)\left({\rm RP}_{\rm iso}(E) + 
                  \sum_{\gamma({\rm iso})} I^{\rm iso}_{\gamma} E^{\rm iso}_{\gamma} \delta(E - E_{\rm iso})\right),
\end{align}
where the first and second sums are over all decaying isotopes and $\gamma$-radiation lines for each isotope, respectively. Each $\gamma$-radiation line is characterized by energy, $E^{\rm iso}_{\gamma}$, and absolute intensity, $I^{\rm iso}_{\gamma}$, per single decay. The spectrum ${\rm RP}_{\rm iso}(E)$ is normalized to unity, ${\int {\rm RP}_{\rm iso}(E)dE = 1}$. Here, we use recent data, provided by the Evaluated Nuclear Reaction Data Library ENDF/B-VIII.0~\footnote{\url{https://www-nds.iaea.org/public/download-endf/ENDF-B-VIII.0/}} library~\citep{brown18}. Finally, $\lambda_{\rm iso}$ is the decay rate of the isotope, and $N_A$ is Avogadro's number. Total \gray{} energy production $\varepsilon_\gamma(t)$ is easily obtained by integrating the spectrum $S(E,t)$ over energy.

While Figure~\ref{fig:gammarates} was computed with a large number of nuclear mass models (25 models), three hydrodynamics conditions for each model and two fission prescriptions for the neutron-rich case, in the rest of the paper we focus on just four representative yield distributions. The nucleosynthesis is computed with parameterized trajectories \citep[an exponential plus power-law decay described in][]{lippuner15} and self-consistent nuclear reheating. Composition of dynamical ejecta is calculated assuming initial entropy ${s = 10\,k_{\rm B}/{\rm baryon}}$, electron fraction ${Y_e = 0.05}$, and expansion timescale ${\tau = 10\ {\rm ms}}$. We explore sensitivity to nuclear physics by using two different fission prescriptions: a symmetric splitting \citep[following][]{mumpower18}, and the fission fragment distributions of~\cite{kodama75}. Composition for both of the "winds" is computed with expansion timescale ${\tau = 100\ {\rm ms}}$, initial entropy ${s = 50\,k_{\rm B}/{\rm baryon}}$, and electron fractions ${Y_e = 0.4}$ for ``wind 1'' and ${Y_e = 0.3}$ for ``wind 2''. This is  similar to the basic compositions used in \cite{wollaeger18}.  

\begin{figure}[htp]
\centering
\begin{tabular}{c}
\includegraphics[width=0.45\textwidth]{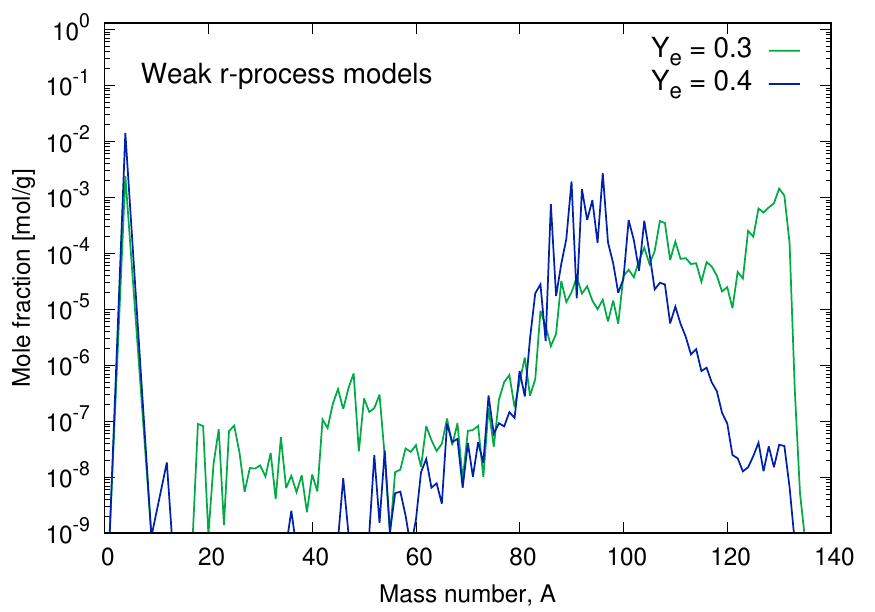} \\
\includegraphics[width=0.45\textwidth]{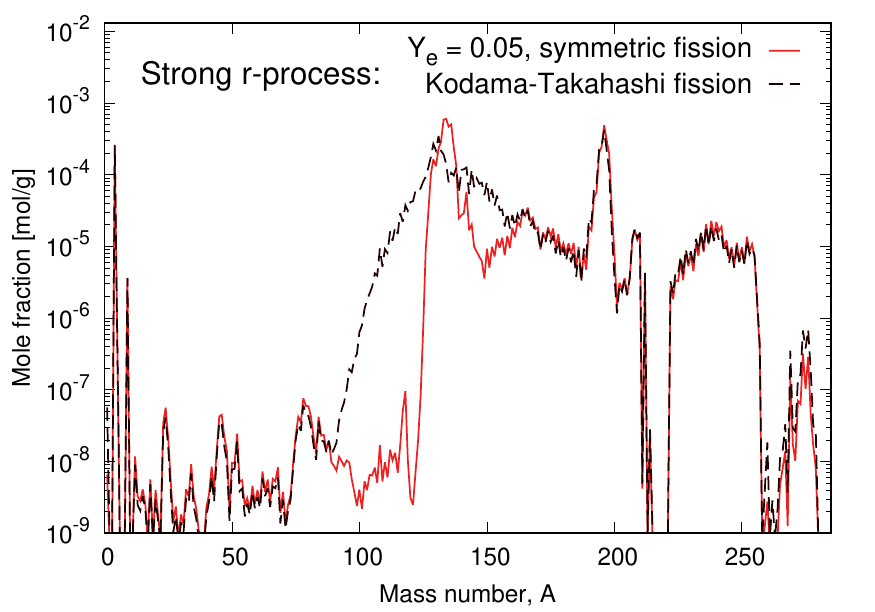}
\end{tabular}
\caption{Model abundances for the weak (top) and strong (bottom) 
$r$-process, sampled at the epoch ${t = 1}$~day.}
\label{fig:abundances}
\end{figure}

\begin{table}
  \caption{Models Summary.} 
\begin{tabular}{lcccc}
\hline
      & Weak        & Strong      & Fission &       \\[-2mm]
Model & $r$-process & $r$-process & Model   & Shape \\[-2mm]
      & (spherical) & (toroidal)  &         & \\
\hline
S1    & $Y_e=0.4$ & ---         & ---        & Spherical \\
S2    & $Y_e=0.3$ & ---         & ---        & Spherical \\
As    & ---       & $Y_e=0.05$  & Symmetric  & Torus \\
Ak    & ---       & $Y_e=0.05$  & KT${}^*$   & Torus \\
S1As  & $Y_e=0.4$ & $Y_e=0.05$  & Symmetric  & Sphere+torus \\
S1Ak  & $Y_e=0.4$ & $Y_e=0.05$  & KT         & Sphere+torus \\
S2As  & $Y_e=0.3$ & $Y_e=0.05$  & Symmetric  & Sphere+torus \\
S2Ak  & $Y_e=0.3$ & $Y_e=0.05$  & KT         & Sphere+torus \\
\hline
\end{tabular}\\

\vspace{5pt}
{\bf Note.}  Columns: model notation; initial $Y_e$ in the 
  high-$Y_e$ outflow (producing weak $r$-process); initial $Y_e$ in the 
  neutron-rich outflow (producing the main / strong $r$-process); fission model;
  and combined shape.\\
${}^*$ "KT" = Kodama-Takahashi fission model (see main text for details).\\
\label{tab:models}
\end{table}

The resulting yield distributions one day after the merger are shown in Figure~\ref{fig:abundances}.  
Two distributions for the low-$Y_e$ dynamical ejecta (red, {\tt As} and black, {\tt Ak}) represent strong $r$-process between the second and third peak, computed with two different types of fission model, as previously described. The medium-$Y_e$ wind component (green, {\tt S2}) spans the range from first to the second $r$-process peaks, while the high-$Y_e$ component (blue, {\tt S1}) only produces the first $r$-process peak. These four uniform-composition models are selected to represent dominant peak contribution. Models {\tt S1} and {\tt S2} have spherically symmetric morphology (``S'') and correspond to the yields with $Y_e=0.4,0.3$, respectively. Models {\tt As} and {\tt Ak} have morphology of model "A" from~\cite{rosswog14} and correspond to the strong $r$-process  production with symmetric split and Kodama--Takahashi fission models, respectively. Superimposing these models, we additionally construct four two-component models. Our models are summarized in Table~\ref{tab:models}.

\section{Gamma Rays from Kilonova Transients}
\label{sec:transients}

The \gray{} emission is strongest at early times (first 10 days) when it emerges from the expanding ejecta. This is the so-called kilonova epoch. For a nearby event, the gravitational-wave and follow-up electromagnetic detections of this event will provide exact localization, allowing dedicated \gray{} follow-up of the kilonova. Initially, most of the emitted gammas are trapped in the flow and the escape of this emission requires transport calculations. The transport necessarily distorts \gray{} source spectrum: it broadens every line and absorbs or redistributes energy. We use the \gray{} production spectra for four representative yields calculated in the previous Section~\ref{sec:nuclear} to source \grays{} in the transport code, and perform 3D transport simulations on a material background of the two morphologies as described above.

For this optically thick transport regime, we use the Monte Carlo \gray{} transport code \texttt{Maverick} described in~\cite{hungerford03,hungerford05}.  In the context of $^{56}$Ni decay in thermonuclear supernovae, this code has been verified in a code comparison effort against most major codes in the community~\citep{milne04}. \texttt{Maverick} assumes the material properties (density and composition) are in steady state for each time slice.  Average escape time of gamma-ray packets is $< 25$\% of the age of the explosion for all time slices considered, so this steady-state assumption is reasonable.  The ejecta is followed assuming a homologous expansion and then mapped into a 3D ($50^3$) grid for the transport. 

We assume that the source spectrum is proportional to the mass in each zone. We use luminosity-weighted packets, so the number of Monte Carlo packets in each zone is also proportional to the mass.  The packets sample the energy of the \grays{} based on our emission spectrum and are binned into 2000 energy groups ranging from 5\,keV to 20\,MeV.

The \gray{} opacity includes components from Compton scattering, photoelectric absorption, and pair production absorption.  It is dominated by Compton scattering above roughly 100-300 keV.  Photoelectric absorption becomes important below 100-300 keV, depending on the $Z$ of the absorbing material.  Compton scattering is treated by sampling the outgoing photon properties (energy and angle) from the complete Klein--Nishina scattering kernel in the free electron limit.  The electron density in each zone is contributed by electrons from the wind component as well as electrons from the dynamic ejecta component.

The photoelectric absorption opacity ($\sigma_{\rm PE}$) is represented as an effective absorption as follows:
\begin{equation}
    \sigma_{\rm PE} = n_{\rm abs} \sigma_{\rm abs} = \rho_{\rm eje}/(m_{\rm proton} \bar{a}) \sigma_{\rm Fe}
\end{equation}
where the number of absorbers ($n_{\rm abs}$) is set to the density of the ejecta ($\rho_{\rm ejecta}$) divided by the average atomic mass ($\bar{A}$) and the proton mass ($m_{\rm proton}$). Here, the ejecta can include both wind and dynamical ejecta components. The cross section per absorber is taken to be the relatively well-known cross section of iron ($\sigma_{\rm Fe}$).  This simplifying assumption for the cross section can lead to errors in our opacity, especially below 100\,keV, where it dominates the opacity as the photoelectric cross section scales as roughly the proton fraction to the fourth power, but it provides a rough estimate for the opacity.  However, features below 100\,keV should be taken with some caution.

\begin{figure*}[htp]
\begin{tabular}{cc}
  \includegraphics[width=0.5\textwidth]{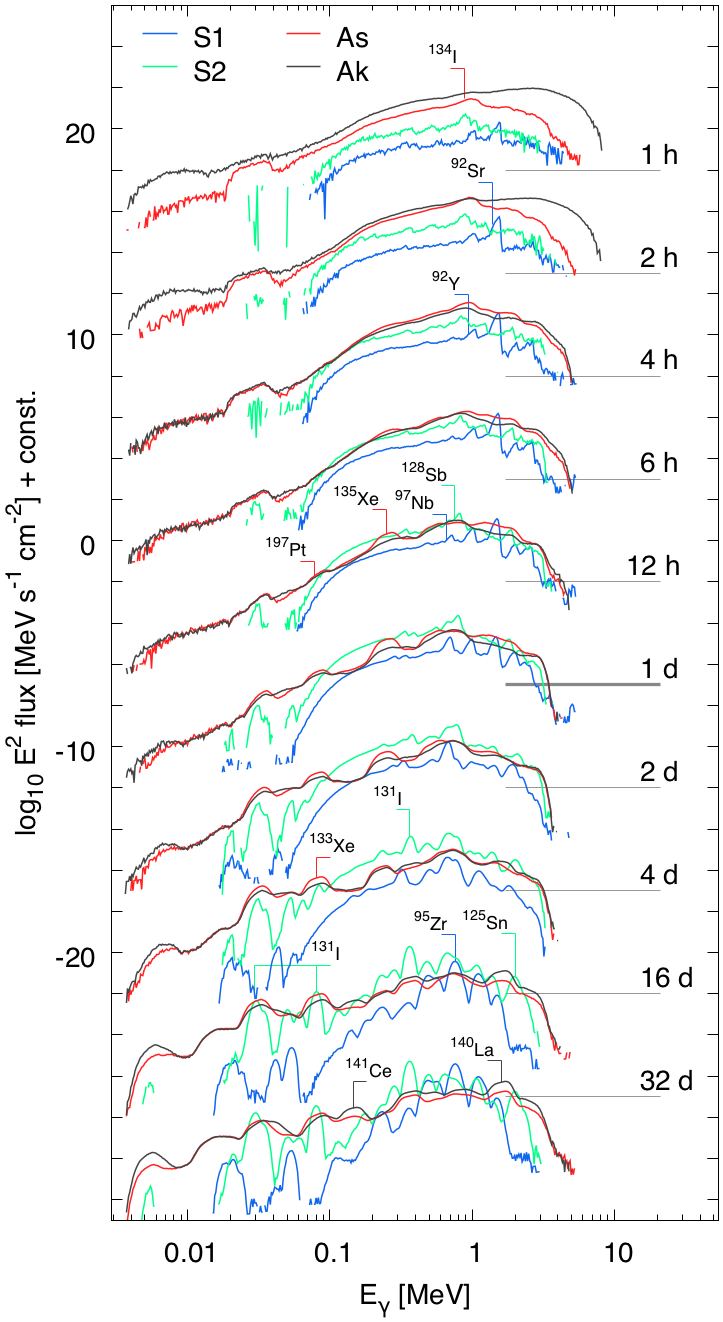} & 
  \includegraphics[width=0.5\textwidth]{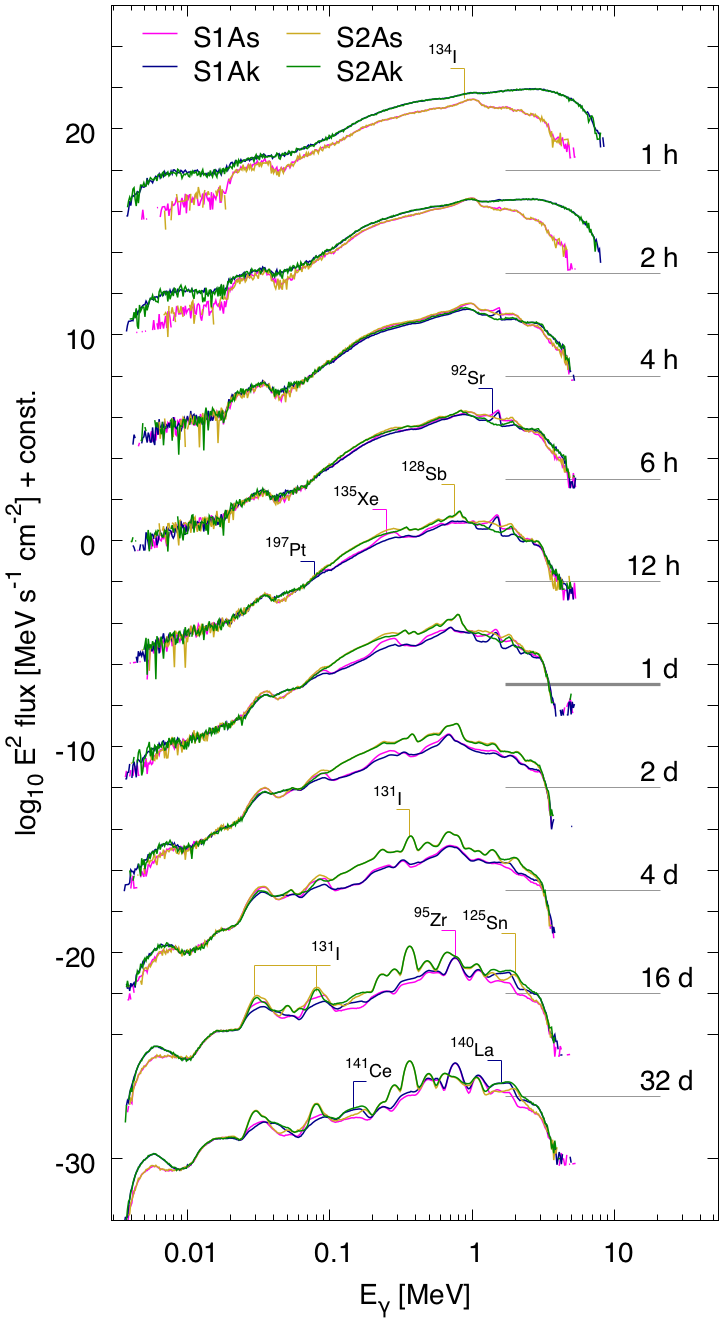}
\end{tabular}  
\caption{Evolution of synthetic spectra of one-component (left) and 
   two-component (right) sources, as seen from the distance of
   3~Mpc. For clarity, the spectra are offset by multiples of 3~dex 
   in log space, up or down from zero-offset spectrum at 1~day. 
   The offsets are indicated by horizontal lines.
   Some of the features in one-component spectra are labeled with 
   isotopes which are producing the features (see Table~\ref{tab:transient}).
} 
\label{fig:transient}
\end{figure*}

With this physics, we use \texttt{Maverick} to calculate the escape fraction and energy of the Monte Carlo packets.  These packets are tallied into a spectrum that has 250 logarithmically spaced energy bins from 3\,keV to 20\,MeV.  Figure~\ref{fig:transient} shows the resulting spectra for both one- and two-component models.  There are distinct differences in the \gray{} signal between all of our models in the first few hours, which persist to late times.  

\begin{table}
    \caption{Some of the Isotopes with Bright Lines which Produce the
    Spectral Peaks Visible in Figure~\ref{fig:transient}.}
    \begin{tabular}{c|c|c|cc} 
        \hline
        Models & Time Range & Line Energy [keV] & Isotope & $T_{1/2}$ \\
        \hline\hline
        \multirow{4}{*}{S1} & 1 hr--1 day   & 1384 & \Sr{92}   & 2.611 hr  \\
                            & 2 hr--2 days  & 934  & \Y{92}    & 3.54 hr   \\
                            & 12 hr--2 days & 658  & \Nb{97}$^*$& 72.1 minutes \\
                            & $>8\;$days    & 756  & \Zr{95}   & 64.0 days   \\
        \hline
        \multirow{6}{*}{S2} & 6 hr--4 days  & 743  & \Sb{128}  & 9.05 hr \\
                            & 6 hr--4 days  & 754  & \Sb{128}  & 9.05 hr \\
                            & $>4\;$days    & 364  & \I{131}   & 8.02 days \\
                            & $>4\;$days    & 80.2 & \I{131}   & 8.02 days \\
                            & $>4\;$days    & 29.8 & \I{131}   & 8.02 days \\
                            & $>4\;$days    & 2002 & \Sn{125}  & 9.64 days \\
        \hline
        \multirow{2}{*}{As, Ak} 
                            & 12 hr--1 day  & 77.4 & \Pt{197}  & 19.9 hr \\
                            & $>2\;$days    & 81.0 & \Xe{133}  & 5.25 days \\
        \hline
        \multirow{3}{*}{As} 
                            & $<6\;$hr      & 847  & \I{134}   & 52.5 minutes \\
                            & $<6\;$hr      & 884  & \I{134}   & 52.5 minutes \\
                            & 12 hr--2 days & 249  & \Xe{135}  & 9.14 hr   \\
        \hline
        \multirow{2}{*}{Ak} 
                            & $>10\;$days   & 145  & \Ce{141}  & 32.5 days   \\
                            & $>10\;$days   & 1596 & \La{140}$^*$& 1.67 days  \\
        \hline
    \end{tabular}
    \\
    
    \vspace{5pt}
    {\bf Note.} Peak energies are listed as the line energy for the responsible isotope.\\
    \footnotesize{(*) Rapidly decaying isotope, continuously produced by a long-lived ancestor}
    \label{tab:transient}
\end{table}

All models show more high-energy spectrum at early times. Models Ak and As lose about one order of magnitude in brightness between the first hour and the first day, while S1 and S2 gain approximately the same amount, catching up and becoming dominant emitters compared to As and Ak around the spectral peak at 1~MeV. Ak model (as well as S2Ak and S1AK) has a distinct enhancement at the first two hours in the range of high \gray{} energies $>5$~MeV. At early times, the line broadening is noticeably blue-shifted due to the photosphere approaching the observer, while at later times (after one day or so) the broadened lines become much more symmetric. Dynamical ejecta models show much less features; this is not so much due to the morphology expanding twice as fast on average as because there are many isotopes contributing and blending to form a pseudo-continuous spectra. Nevertheless, lines from certain radioactive nuclides such as \Pt{197} and \Xe{133} can be clearly identified. Isotopes \La{140} and \Ce{141} are only prominent for the Ak model, while \Xe{135} and \I{134} emerge for the As model, for which a different fission prescription was used. This gives a hint that potentially a correct fission model can be decided from the observation. Table~\ref{tab:transient} lists the properties of the most prominent \gray{} producing isotopes for each of the models.

An important feature distinguishing kilonova transients in \grays{} is that the lines of individual decaying nuclides become prominent only on the timescale comparable to their mean lifetime. This is unlike the optical or infrared signal, which is affected by the entire yield at all times. Bright \gray{} emitting isotopes take turns to emerge in the spectrum, allowing the potential to trace evolution of the composition in real time. However, this effect is mitigated by the long integration time, even for the most sensitive detectors.

In summary, \gray{} observations will be able to determine whether the ejecta originates from the electron poor or electron rich initial conditions. However, the differences between fission models Ak versus As are very small and will be difficult to detect. In models with mixed electron fractions and multiple components, it will be difficult to determine the exact yield (only that the material is mixed and not dominated by a low- or high-electron fraction abundance.  After 10~days, the emission has dropped by two orders of magnitude, becoming increasingly difficult to detect.

\section{Gamma Rays from Kilonova Remnants}
\label{sec:remnants}

The detection of old kilonova remnants provides an alternate observational prospect to constraining the nucleosynthesis in neutron star mergers. Although the rate of neutron star mergers is about three orders of magnitude lower that that of supernovae, given the fact that a few hundred supernova remnants have been discovered, it is not unreasonable to assume that kilonova remnants younger than 100\,kyr can be found in our neighborhood of the Milky Way~\citep{wu19}. If a relatively young remnant exists close to the Earth, we may be able to detect it and probe the yields of the merger. The \gray{} spectrum of a kilonova remnant would consist of multiple lines generated by long-lived residual nuclides from the $r$-process, providing unique perspective on its nuclear physics. As the remnant decelerates, line broadening is less important~\citep{piran13}, producing individually identifiable lines of specific radionuclides. This can be particularly helpful for discriminating between various $r$-process scenarios. This is true even for very dilute interstellar medium in the Galactic halo. In this section, we study both the remnant evolution to determine velocities and spatial sizes of kilonova remnants and the expected \gray{} signals, comparing the results from two fission models.

\subsection{Kilonova Remnant Evolution and Properties}

An explosive remnant (whether it be a supernova or kilonova) passes through four evolutionary phases:  free expansion, Sedov--Taylor, snowplow and merger with interstellar medium. The free expansion phase is assumed to last until the ejecta sweeps up mass comparable to itself. During this phase, the expectation is that the ejecta is expanding without decelerating. The velocity of the shock ($v_{\rm shock}$) is a constant and the radius of the shock ($r_{\rm shock}$) increases with time ($t$) linearly.

When radiative cooling is slow compared with the shock evolution, the Sedov--Taylor similarity solution~\citep{taylor41,taylor50,sedov46} is used to model the shock evolution.  This similarity solution can be derived through simple dimensional analysis:  $[E_{\rm exp}/\rho_{\rm CSM}]$ have units of $({\rm g}\;{\rm cm}^2\;{\rm s}^{-2})/({\rm g}\;{\rm cm}^{-3}) = {\rm cm}^5\; {\rm s}^{-2}$.  With these units, we can derive the shock position,
\begin{equation}
    r_{\rm shock}=(E_{\rm exp}/\rho_{\rm CSM})^{2/5} t^{2/5}
\end{equation}
where $E_{\rm exp}$ is the explosion energy, $\rho_{\rm CSM}$ is the density of the circumstellar medium which is, for massive stars, the stellar wind, and for neutron stars, the interstellar medium.  For a blast wave moving through a constant-density medium, the radius increases as time to the $2/5$ power.  The corresponding shock velocity ($\rm v_{\rm shock}$) is
\begin{equation}
    v_{\rm shock} = d r_{\rm shock}/dt \approx (E_{\rm exp}/\rho_{\rm CSM})^{2/5} t^{-3/5}.
\end{equation}

This phase continues until radiative cooling becomes faster than the evolution of the shock.  At this point, the shock evolves through a snowplow phase where the evolution is dictated by momentum conservation.  In this phase, the remnant velocity ($v_{\rm shock}$) is
\begin{equation}
    v_{\rm shock} = v_{\rm ejecta} m_{\rm ejecta}/(m_{\rm ejecta} + 4 \pi r_{\rm shock}^3 \rho_{\rm CSM}) 
\end{equation}
where $m_{\rm ejecta}$ is the ejecta mass and $v_{\rm ejecta}$ is the ejecta velocity.  At late times, the ejecta mass can be neglected in the denominator and the radius as a function of time is
\begin{equation}
    r_{\rm shock}=(v_{\rm ejecta} m_{\rm ejecta}/\pi \rho_{\rm CSM})^{1/4} t^{1/4}.
\end{equation}

To determine how well these simple analytic estimates match the properties of the remnant, we have modeled the ejecta expansion numerically in 1D to late times.  For the purposes of this study, two properties are most crucial:  the velocity distribution of the radioactive ejecta to get line broadening and the extent of the remnant.  Our numerical model uses a 1D Lagrangian hydrodynamics code initially designed for supernovae~\citep{fryer99c}, but modified (using a simple $\gamma=5/3$ equation of state) to follow the ejecta out to large distances.  With this code, we calculate several models with varying ejecta masses, velocities and densities of the circumstellar medium.  Because of the kicks imparted on neutron stars at birth, the merger can happen far off of the Galactic plane where the density of the surrounding medium is low, spanning a large range: ${10^{-4}-10^2\;{\rm cm^{-3}}}$ \citep{wiggins18}.  

\begin{figure}[!hbtp]
\centering
\includegraphics[width=\columnwidth]{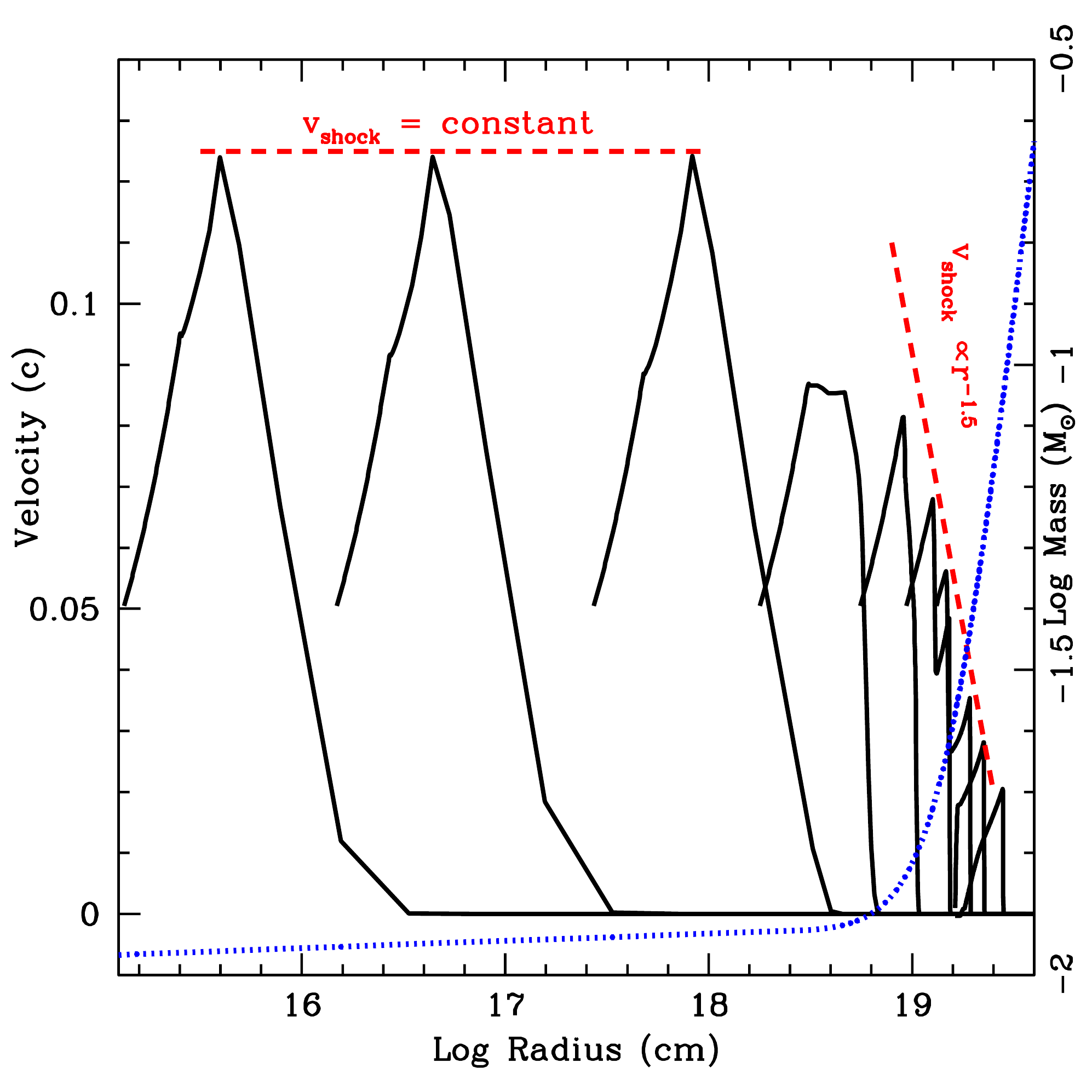}
\caption{Profile of the remnant expansion velocity at a series of times after the explosion, from the free-streaming phase to the Sedov phase.  In general, the simple analytic solutions (dashed red lines) match the numerical hydrodynamic solutions, but there is a transition region that is not exactly fit by the simple solutions.  Nonetheless, for the estimates made here, it is clear that the analytic solutions are a reasonable estimate.  A reverse shock is produced in these calculations that will heat the ejecta, possibly leading to X-ray and radio emission.  The transition region also marks the time when the remnant starts to sweep significant amount of mass from the interstellar medium (dotted purple curve).}
\label{fig:evol}
\end{figure}

Before we discuss the full suite of results, consider the evolution of the explosion better.  Figure~\ref{fig:evol} shows the velocity profile of a shock from a kilonova explosion with $0.01\,M_\odot$ of ejecta, $6\times10^{49}$\,erg of energy, and an interstellar medium (ISM) density of roughly $0.001\,cm^{-3}$ for times ranging from 10~days to 500~yr.  It takes over 100~yr for the shock to sweep up a mass equal to the ejecta mass fully transition to the Sedov--Taylor phase where the velocity decreases with the radius to the 3/2 power.   Note that there is a transition region where the shock decelerates but not as quickly as expected with Sedov--Taylor. 

\begin{figure}[!hbtp]
\centering
\includegraphics[width=\columnwidth]{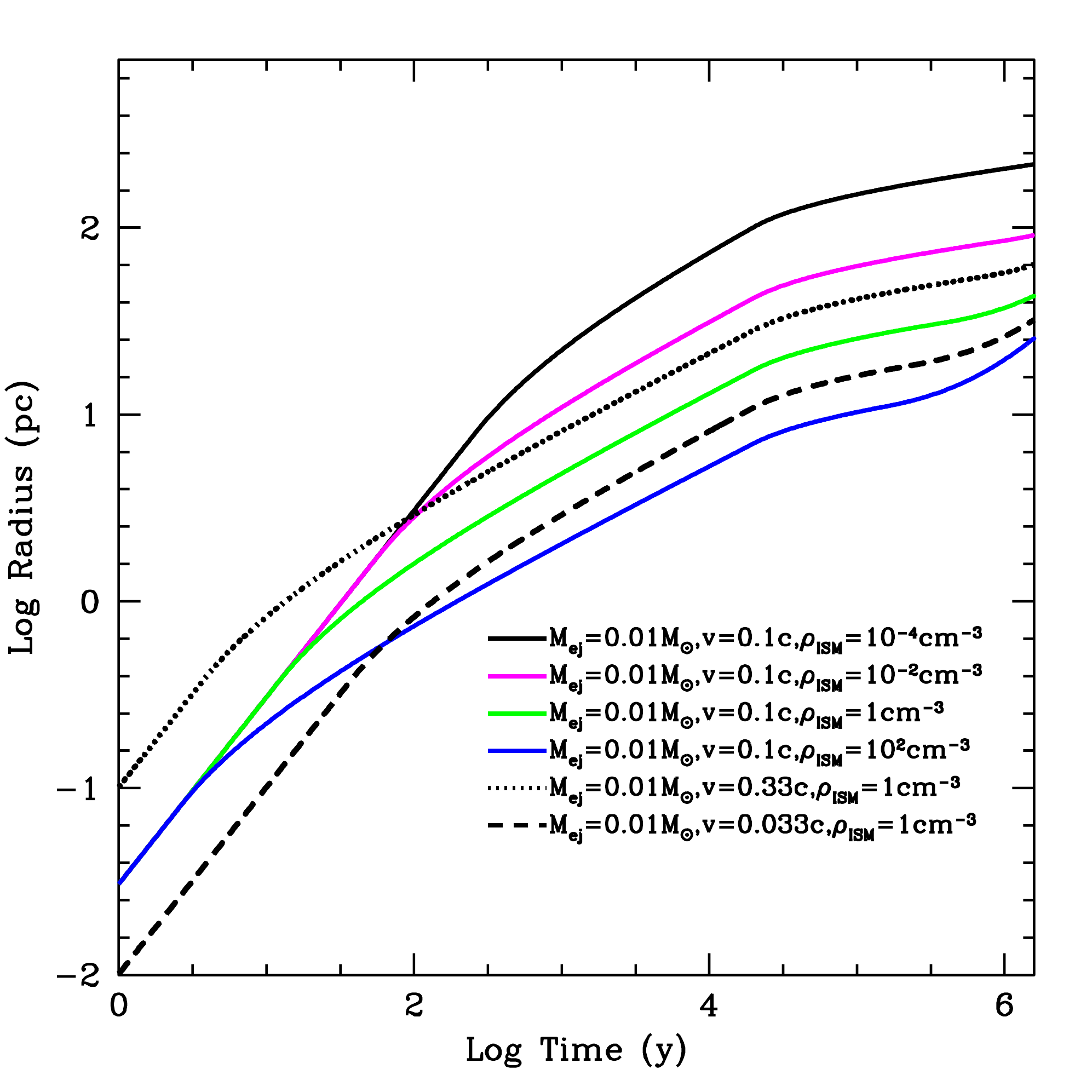} \\
\includegraphics[width=\columnwidth]{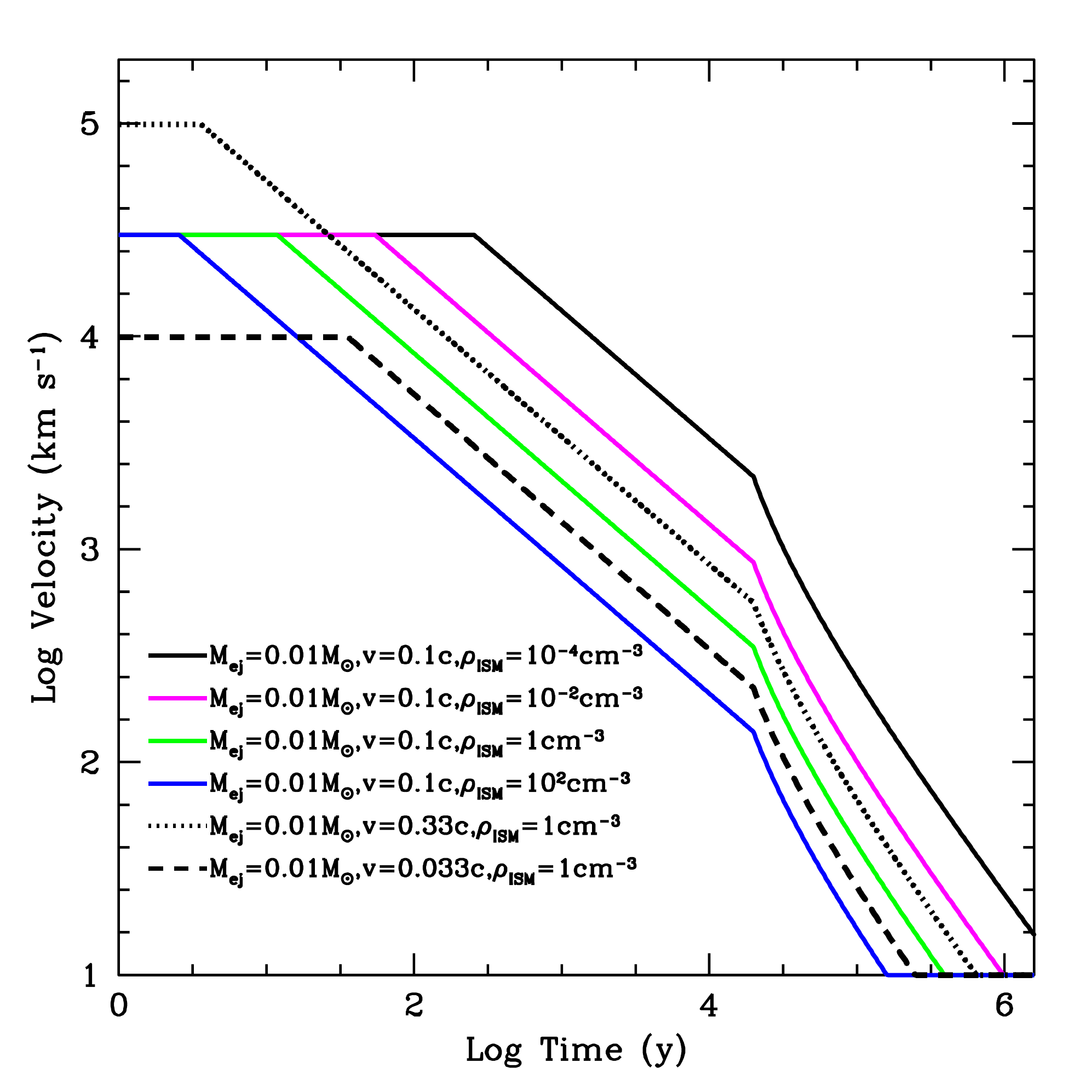}
\caption{ Top: size of the kilonova remnant as a function of time for a range of kilonova ejecta and interstellar medium (ISM) properties.  Kilonovae are typically faster than supernovae, but have less mass (so decelerate more quickly).  Neutron star mergers are expected to occur in lower densities~\citep{wiggins18} than supernovae and, in some cases, these remnants can expand more dramatically than their supernova counterparts.
Bottom: velocity of the kilonova remnant forward shock radius as a function of time for a range of kilonova ejecta and interstellar medium (ISM) properties as in figure~\ref{fig:rvst}.
} 
\label{fig:rvst}
\end{figure}

We have constructed models of the kilonova remnant, coupling the four phases of the remnant evolution to determine both the remnant size and velocity (Fig.~\ref{fig:rvst}) as a function of time. Within a factor of two or so, the late-time properties of these remnants ($>10^4$ yr) are not very different from supernovae.  Although the velocities are higher in kilonovae, the lower ejecta masses mean that the kilonova remnant decelerates faster than normal supernovae.  We also expect radiative cooling to dominate sooner when the kilonova remnant transitions from the Sedov to the snowplow phase at earlier times, leading to more rapid deceleration after roughly 10,000\,yr.  At $10^4$\,yr, the kilonova remnant is expanding at between one hundred and a few thousand ${\rm km \, s^{-1}}$ and at $10^5$\,yr may already have decelerated to the sound speed of the ISM (tens of ${\rm km \, s^{-1}}$) and still expanding at one hundred ${\rm km \, s^{-1}}$.

An interesting feature of kilonova remnants is the rapid evolution to the Sedov phase. Whereas supernova remnants are free streaming for the first 100--1,000\,yr (depending on the density of the interstellar medium), kilonovae enter this phase between 0.25 and 100\,yr.  During the Sedov phase, a reverse shock is produced that heats the ejecta, driving strong radio emission, detectable sometimes within a year or a few years from the outburst~\citep{piran13}.

\subsection{Remnant Gamma Rays}
\label{sec:knrgr}

It is estimated that a few neutron star merger remnants exist in our neighborhood of the Milky Way with ages below a 100\,ky \citep{ripley14,wu19}. As follows from the previous section, for remnants with 10-100\,kyr ages, the ejecta velocities are likely to lie between 100 and $3,000 {\rm km \, s^{-1}}$ and the remnant size lies between 5 and 300\,pc. A remnant 3,kpc away from the Earth will have angular size $0.3^\circ-6^\circ$. In this section, we review the expected \gray{} spectra from these remnants as a function of composition. In the same way as atomic spectra can be used to infer composition, these decay spectra can be used as fingerprints of the yields. At these ages, velocity broadening will not significantly alter the line signals.

\begin{figure*}[htp]
\begin{tabular}{cc}
 \includegraphics[width=0.5\textwidth]{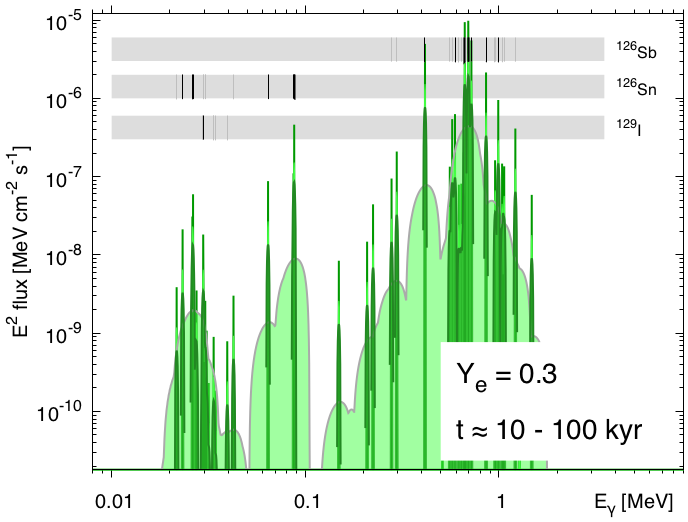} &
 \includegraphics[width=0.5\textwidth]{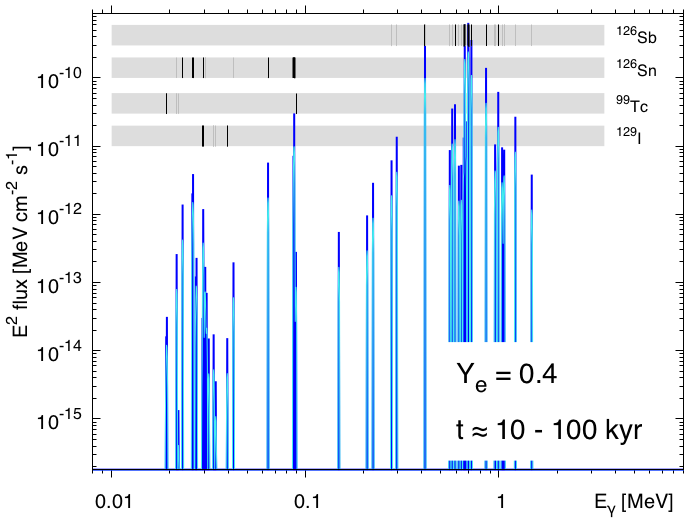}
\end{tabular}
\caption{Gamma-ray spectra of the outflows with moderate neutron richness for the period 
${t\approx10-100\;{\rm kyr}}$ broadened with expansion velocities 
${100 - 3000\;{\rm km}\;{\rm s}^{-1}}$.
Left panel: outflow with $Y_e=0.3$; right column: neutron-poor outflow with $Y_e=0.4$. 
Mass of each outflow: ${m=0.01\;M_\odot}$. 
Distance to the source: ${D = 10\;{\rm kpc}}$.
Dark- and light-shaded spectra are broadened to $1\%$ and $10\%$, respectively, emulating spectral sensitivity of the detector.
} 
\label{fig:latetimespec}
\end{figure*}

For neutron-poor ejecta with a composition peaking near the first $r$-process peak, many of the isotopes have already decayed by 10--100\,kyr.  But a few isotopes, $^{99}$Tc, $^{126}$Sn, $^{126}$Sb, and $^{129}$I, contribute to the \gray{} spectra with energy spanning from roughly 30\,eV to a few MeV.   Figure~\ref{fig:latetimespec} shows the \grays{} for both $Y_e=0.3$ and $Y_e=0.4$ ejecta.  The decay timescales for these isotopes are long (more than 100\,kyr) and the signal at 10\,kyr is not so different than the signal at 100\,kyr.

\begin{figure}[h]
\centering
\includegraphics[width=3.0in]{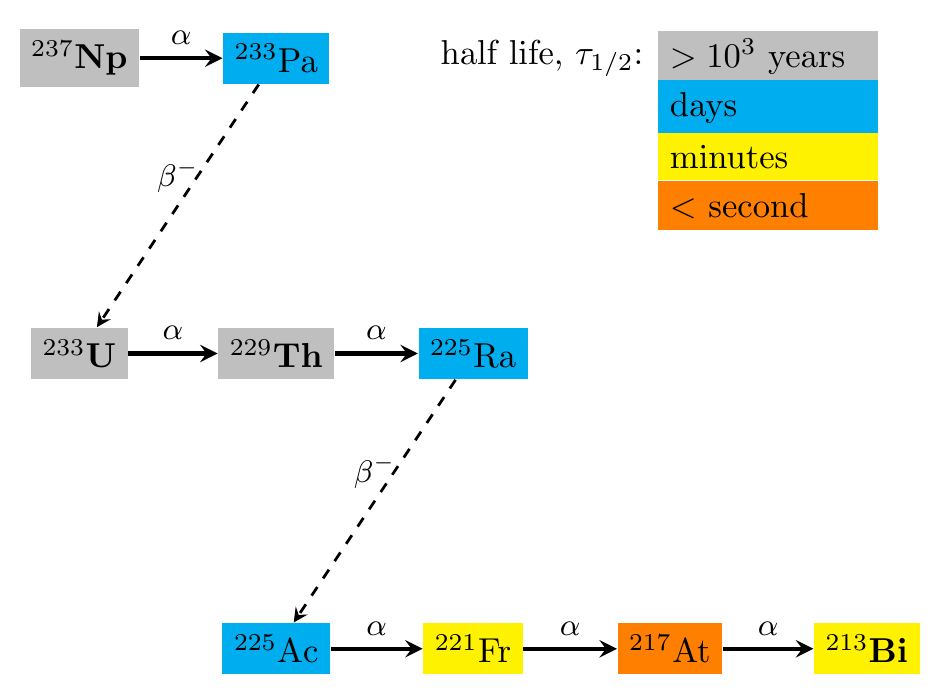}
\caption{Complex decay chain responsible for the production of $^{213}$Bi, 
which is a potentially detectable \gray{} emitter. With the half-life of about 
45 minutes, its presence in neutron star merger remnants can only indicate large 
quantities of one of the long-lived ancestor isotopes: $^{229}$Th (half-life 
7880 yr), $^{233}$U (${1.592\times10^5}$ yr) or, on longer time scales, 
$^{237}$Np ($2.14\times10^6$ yr).}
\label{fig:dchain}
\end{figure}

If we instead focus on the neutron-rich ejecta, both the spectra and the physics are more involved, as complex decay pathways may arise leading to nonintuitive \gray{} emitters. Figure \ref{fig:dchain} shows one such example where $^{237}$Np, the long-lived ancestor with $T_{1/2}\sim2.1 \times 10^{6}$ yr, decays into \gray{} producing $^{213}$Bi, a nucleus whose half-life is roughly 45 minutes. 

\begin{figure*}[htp]
\begin{tabular}{cc}
\includegraphics[width=0.5\textwidth]{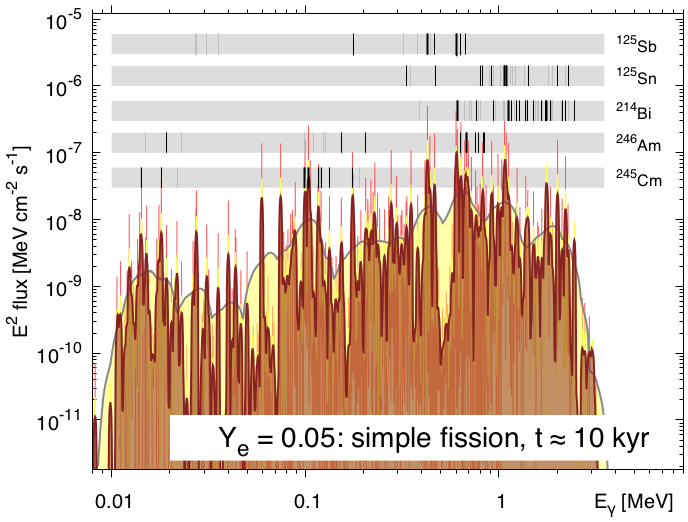} &
\includegraphics[width=0.5\textwidth]{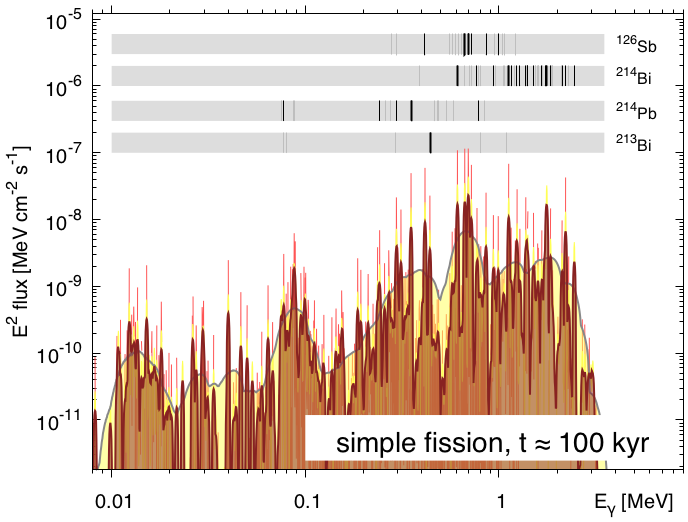}
\\
\includegraphics[width=0.5\textwidth]{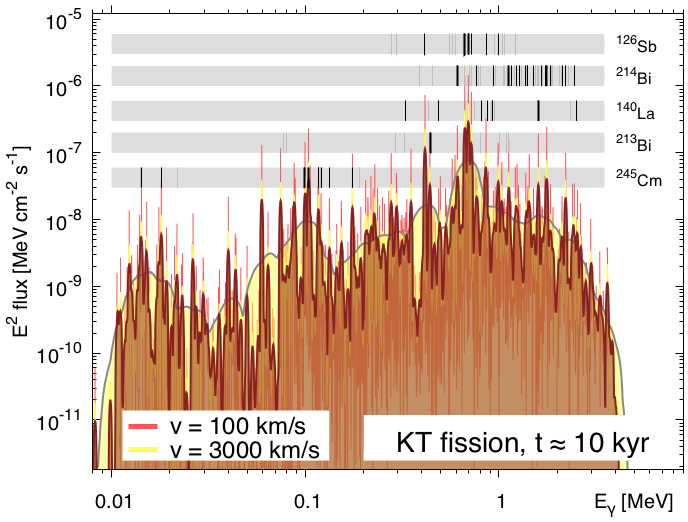} &
\includegraphics[width=0.5\textwidth]{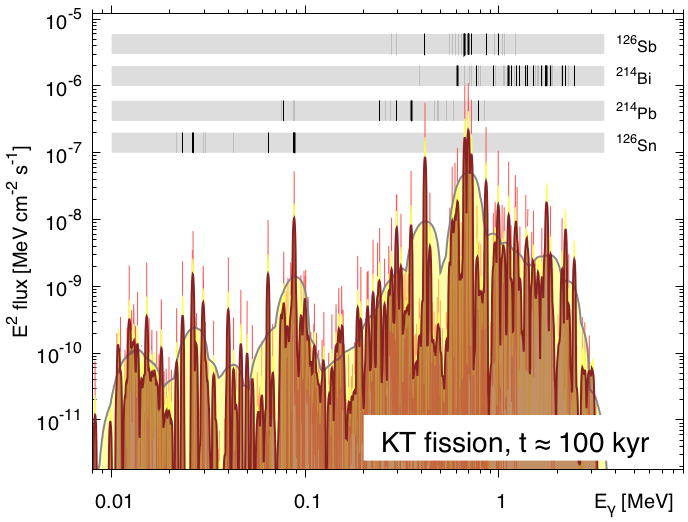}
\end{tabular}
\caption{Broadened \gray{} spectra of a neutron-rich ("red") remnant at 10~kyr (left panels) and 100~kyr (right panels) after the merger. Top: symmetric-split fission model; bottom: Kodama--Takahashi fission product distribution. 
Dark- and light-shaded spectra are broadened to  $1\%$ and $10\%$, respectively, emulating spectral sensitivity of the detector.
} 
\label{fig:latetimespec_red}
\end{figure*}

At sufficiently high neutron richness, the yields are less sensitive to the exact neutron fraction, but more sensitive to the nuclear physics such as the fission model, reinforcing the need for improved nuclear physics modeling for the $r$-process \citep{horowitz18}. Figure~\ref{fig:latetimespec_red} shows the spectra at 10 and 100\,kyr for our dynamical ejecta with two different fission models. As with the atomic spectra, there is a forest of decay lines. Because of the forest of lines, velocity broadening can merge lines and we include plots at the low and high end (100, 3000\,km\,s$^{-1}$) of our remnant velocities. With expected energy resolutions, it may still be possible to distinguish between the yields of different nuclear physics models. 

Network calculations of neutron-rich ejecta suggest $^{126}$Sb, $^{128}$Sb, $^{214}$Bi, $^{214}$Pb, $^{243}$Am, $^{246}$Am, $^{245}$Cm, and $^{250}$Bk are the dominant isotopes contributing to the spectra on the observational timescale of 10 and 100\,kyr. These isotopes are the result of decays from long-lived ancestors that set the observational timescale. We summarize possible influential \gray{} emitters and their long-lived ancestors in Tables \ref{tab:gammas} and \ref{tab:100ky_gammas}. Some of these isotopes have been studied as potential indicators of the $r$-process in previous works~\cite{qian98,ripley14,wu19} as they can be found by surveying the half-lives in the nuclear chart. Other isotopes emerge as prominent emitters despite having short half-lives as decay products downstream of long-lived populating ancestors. This demonstrates how comprehensive treatment of nucleosynthesis is significant when catching \gray{} emitters.

Fission of actinides produces several short-lived isotopes which leave distinct mark on the \gray{} spectrum. Table~\ref{tab:gammas} lists two such isotopes, \I{134} and \La{140} being signature for the symmetric-split fission model, and two other, \Sb{125} and \Sn{125} as signature isotopes for the Kodama--Takahashi fission. Both Tables~\ref{tab:gammas} and \ref{tab:100ky_gammas} list \Sn{126} as an isotope which can be either synthesized from the very beginning, or as a signature fission product of the Kodama--Takahashi fission prescription. If a remnant were to be discovered nearby, this hints to the exciting possibility to observe \gray{} lines of short-lived fragments of actinide fission, such as those actively studied in experiment, e.g., the CAlifornium Rare Isotope Breeder Upgrade (CARIBU) facility at Argonne National Laboratory~\citep{marley13,vanschelt13}. Observing such lines will allow to reason about models of nuclear fission.

Tables \ref{tab:gammas} and \ref{tab:100ky_gammas} show that for a remnant at distance of 3~kpc away none of the isotopes produce line flux in excess of {$10^{-6}$ photons s$^{-1}$ cm$^{-2}$}. Sensitivity of at least {$10^{-7}$ photons s$^{-1}$ cm$^{-2}$} is required to detect such a remnant. For a remnant that is 100\,kyr old, the line flux is an order of magnitude less. \cite{ripley14} suggested blind search for neutron star merger remnants within the Galactic plane. However, an older remnant is possible in the halo, away from the Galactic plane. At 100\,kyr in the halo with low density of interstellar medium the remnant would have the radius of $60-200$\,pc (see Figure~\ref{fig:rvst}) and the angular size of $1^\circ-6^\circ$~\citep[see also][]{wu19}.

\begin{table*}
\begin{adjustwidth}{-.15in}{-.15in}
\centering
\footnotesize
  \caption{Possible Influential \gray{} Emitters and their Long-lived Populating Ancestors as Found by Network Calculations on a 10 ky Observational Timescale.
  } 
\begin{tabular}{ccc|ccc|cc}
\hline
Isotope & $T_{1/2}$ & Mass Range [M$_\odot$] & Ancestor(s) & $T_{1/2}$ & Ancestor Mass Range [M$_\odot$] & Line Energy [keV] & Flux [ph s$^{-1}$ cm$^{-2}$] \\
\hline\hline
\Am{241} & 432.6 yr & (1 -- 10)$\times 10^{-9}$ & {\Cm{245}} & {8423 yr} & {$(2 - 20)\times 10^{-8}$} & \textit{\textbf{59.5409}} & $(2 - 20)\times 10^{-8}$ \\
\hline
{\Am{243}} & {7364 yr} & {(1 -- 10)$\times 10^{-8}$} & {Self} & & & \textit{\textbf{74.66}} & $(2 - 30)\times 10^{-8}$ \\
\hline
\multirow{2}{*}{\Am{246}} & \multirow{2}{*}{39 minutes} & \multirow{2}{*}{(1 -- 10)$\times 10^{-17}$} & \multirow{2}{*}{\Cm{250}} & \multirow{2}{*}{8300 yr} & \multirow{2}{*}{$(8 - 80)\times 10^{-9}$} & 679.2 & $(3 - 30)\times 10^{-9}$ \\
 & & & & & & 756 & $(7 - 70)\times 10^{-10}$ \\
 \hline
{\Bi{213}} & {45.59 minutes} & {$(5 - 60)\times 10^{-17}$} & {\Th{229}} & {7880 yr} & {$(5 - 50)\times 10^{-9}$} & \textit{440.45} & $(4 - 40)\times 10^{-9}$ \\
\hline
\multirow{3}{*}{\Bi{214}} & \multirow{3}{*}{19.9 minutes} & \multirow{3}{*}{(1 -- 10)$\times 10^{-17}$} & \Ra{226} & 1600 yr & $(5 - 50)\times 10^{-10}$ & 609.32 & $(3 - 40)\times 10^{-9}$ \\
 & & & \Th{230} & 75.4 ky & $(2 - 20)\times 10^{-8}$ & 1120.294 & (1 -- 10)$\times 10^{-9}$ \\
 & & & & & & 1764.491 & (1 -- 10)$\times 10^{-9}$ \\
\hline
{\Bk{250}} & {3.21 hr} & {$(3 - 30)\times 10^{-17}$} & {\Cm{250}} & {8300 yr} & {$(8 - 80)\times 10^{-9}$} & 1028.654 & (1 -- 10)$\times 10^{-10}$ \\
\hline
\multirow{4}{*}{\Cm{245}} & \multirow{4}{*}{8423 yr} & \multirow{4}{*}{$(2 - 20)\times 10^{-8}$} & \multirow{4}{*}{Self} & & & \textit{\textbf{99.5232}} & (1 -- 10)$\times 10^{-8}$ \\
 & & & & & & \textit{\textbf{103.741}} & $(2 - 20)\times 10^{-8}$ \\
 & & & & & & 117.2322 & $(4 - 40)\times 10^{-9}$ \\
 & & & & & & \textit{175.01} & $(5 - 50)\times 10^{-9}$ \\
 \hline
\multirow{3}{*}{\I{134}} & \multirow{3}{*}{52.5 minutes} & \multirow{3}{*}{$<3\times 10^{-17}$} & \multirow{3}{*}{K--T fission} & \multirow{3}{*}{} & \multirow{3}{*}{} 
           & 847.025 & $<8\times 10^{-9}$ \\
 & & & & & & 884.09  & $<5\times 10^{-9}$ \\
 & & & & & & 1072.55 & $<1\times 10^{-9}$ \\
\hline
\multirow{2}{*}{\La{140}} & \multirow{2}{*}{1.68 days} & \multirow{2}{*}{$<2\times 10^{-15}$} & \multirow{2}{*}{K--T fission} & \multirow{2}{*}{} & \multirow{2}{*}{} & 487.021 & $<5\times 10^{-9}$ \\
 & & & & & & 1596.21 & $<10^{-8}$ \\
\hline
\multirow{4}{*}{\Np{239}} & \multirow{4}{*}{2.36 days} & \multirow{4}{*}{(1 -- 10)$\times 10^{-14}$} & \multirow{4}{*}{\Am{243}} & \multirow{4}{*}{7364 yr} & \multirow{4}{*}{(1 -- 10)$\times 10^{-8}$} & \textit{99.5232} & $(4 - 50)\times 10^{-9}$ \\
 & & & & & & \textit{103.741} & $(7 - 80)\times 10^{-9}$  \\
 & & & & & & \textit{\textbf{106.123}} & $(8 - 90)\times 10^{-9}$ \\
 & & & & & & \textit{277.599} & $(5 - 50)\times 10^{-9}$ \\
\hline
\multirow{3}{*}{\Pb{214}} & \multirow{3}{*}{27.06 minutes} & \multirow{3}{*}{$(2 - 20)\times 10^{-17}$} & \Ra{226} & 1600 yr & $(5 - 50)\times 10^{-10}$ & 241.995 & $(5 - 60)\times 10^{-10}$ \\
 & & & \Th{230} & 75.4 ky & $(2 - 20)\times 10^{-8}$ & 295.2228 & (1 -- 10)$\times 10^{-9}$ \\
 & & & & & & 351.9321 & $(3 - 30)\times 10^{-9}$ \\
\hline
\multirow{4}{*}{\Sb{125}} & \multirow{4}{*}{2.76 yr} & \multirow{4}{*}{$10^{-18}$ -- $10^{-11}$} & \multirow{4}{*}{Symm fission} & \multirow{4}{*}{} & \multirow{4}{*}{} & \textbf{427.874} & $8\times 10^{-15}$ -- $10^{-7}$ \\
 & & & & & & 463.365 & $3\times 10^{-15}$ -- $4\times 10^{-8}$ \\
 & & & & & & 600.597 & $10^{-15}$ -- $10^{-7}$ \\
 & & & & & & 635.95  & $10^{-15}$ -- $10^{-8}$ \\
\hline
\multirow{4}{*}{\Sb{126}} & \multirow{4}{*}{12.35 days} & \multirow{4}{*}{$(.3 - 30)\times 10^{-14}$} & \multirow{4}{*}{\Sn{126}} & \multirow{4}{*}{230 ky} & \multirow{4}{*}{$(.2 - 20)\times 10^{-7}$} & \textbf{414.7} & $(.2 - 20)\times 10^{-8}$ \\
 & & & & & & \textbf{666.5} & $(.2 - 20)\times 10^{-8}$ \\
 & & & & & & \textbf{695.0}   & $(.2 - 20)\times 10^{-8}$ \\
 & & & & & & \textbf{720.7} & (.1 -- 10)$\times 10^{-8}$ \\
\hline
\multirow{4}{*}{\Sn{125}} & \multirow{4}{*}{9.64 days} & \multirow{4}{*}{$<4\times 10^{-13}$} & \multirow{4}{*}{Symm fission} & \multirow{4}{*}{} & \multirow{4}{*}{} & 822.48 & $<2\times 10^{-8}$ \\
 & & & & & & 915.55 & $<2\times 10^{-8}$ \\
 & & & & & & 1067.1 & $<4\times 10^{-8}$ \\
 & & & & & & 1089.15 & $<2\times 10^{-8}$ \\
\hline
\multirow{3}{*}{\Sn{126}} & \multirow{3}{*}{230 ky} & \multirow{3}{*}{$(.2 - 20)\times 10^{-7}$} & \multirow{3}{*}{K--T, Self} & \multirow{3}{*}{} & \multirow{3}{*}{} & 64.281 & $(.2 - 20)\times 10^{-9}$ \\
 & & & & & & 86.938 & $(.2 - 20)\times 10^{-9}$ \\
 & & & & & & 87.567 & $(.7 - 70)\times 10^{-9}$ \\
\hline
\end{tabular}
\\
    
\vspace{5pt}
\begin{flushleft}
{\bf Note.} Where the ancestors are fissioning heavy nuclei, the most productive fission yield model is indicated. The half-life $T_{1/2}$ and computed quantity in solar masses are shown for each isotope.  Photon fluxes are computed for a remnant at 3 kpc with total merger ejecta between 0.002 and 0.02 M$_\odot$.  The top ten lines for the minimum flux estimate are shown in \textit{italics}, and the top ten for the maximum flux estimate are shown in \textbf{boldface}.
\end{flushleft}
\label{tab:gammas}
\end{adjustwidth}
\end{table*}

\begin{table*}
\begin{adjustwidth}{-.15in}{-.15in}
\centering
\footnotesize
  \caption{Same as Table \ref{tab:gammas}, Except with a 100 ky Observational Timescale.  
  } 
\begin{tabular}{ccc|ccc|cc}
\hline
Isotope & $T_{1/2}$ & Mass Range [M$_\odot$] & Ancestor(s) & $T_{1/2}$ & Ancestor Mass Range [M$_\odot$] & Line Energy [keV] & Flux [ph s$^{-1}$ cm$^{-2}$] \\
\hline\hline
{\Am{243}} & {7364 yr} & {(1 -- 10)$\times 10^{-9}$} & {\Cm{247}} & {15.6 My} & {(4 -- 40)$\times 10^{-9}$} & \textit{\textbf{74.66}} & (3 -- 30)$\times 10^{-9}$ \\
\hline
{\Bi{213}} & {45.59 minutes} & {(7 -- 70)$\times 10^{-18}$} & {\U{233}} & {159.2 ky} & {(1 -- 10)$\times 10^{-8}$} & 440.45 & (5 -- 50)$\times 10^{-10}$ \\
\hline
\multirow{4}{*}{\Bi{214}} & \multirow{4}{*}{19.9 minutes} & \multirow{4}{*}{(6 -- 60)$\times 10^{-18}$} & & & & \textit{\textbf{609.32}} & (2 -- 20)$\times 10^{-9}$ \\
 & & & {\Th{230}} & {75.4 ky} & {(1 -- 10)$\times 10^{-8}$} & 1120.294 & (5 -- 60)$\times 10^{-10}$ \\
 & & & {\U{234}} & {245.5 ky} & {(2 -- 20)$\times 10^{-8}$} & 1238.122 & (2 -- 20)$\times 10^{-10}$ \\
 & & & & & & 1764.491 & (5 -- 60)$\times 10^{-10}$ \\
\hline
\multirow{4}{*}{\Np{239}} & \multirow{4}{*}{2.36 days} & \multirow{4}{*}{(1 -- 10)$\times 10^{-15}$} & & & & 99.5232 & (6 -- 60)$\times 10^{-10}$ \\
 & & & {\Am{243}} & {7364 yr}  & $< 10^{-10}$              & \textit{103.741} & (6 -- 60)$\times 10^{-10}$ \\
 & & & {\Cm{247}} & {15.6 My} & (2 -- 20)$\times 10^{-8}$ & \textit{106.123} & (6 -- 60)$\times 10^{-10}$ \\
& & & & & & 277.599 & (6 -- 70)$\times 10^{-10}$ \\
\hline
\multirow{3}{*}{\Pa{233}} & \multirow{3}{*}{26.98 days} & \multirow{3}{*}{(3 -- 30)$\times 10^{-15}$} & \multirow{3}{*}{\Np{237}} & \multirow{3}{*}{2.144 My} & \multirow{3}{*}{(8 -- 90)$\times 10^{-8}$} & 300.129 & (6 -- 70)$\times 10^{-11}$ \\
 & & & & & & 311.904 & (4 -- 40)$\times 10^{-10}$ \\
 & & & & & & 340.476 & (4 -- 50)$\times 10^{-11}$ \\
\hline
\multirow{3}{*}{\Pb{214}} & \multirow{3}{*}{27.06 minutes} & \multirow{3}{*}{(8 -- 80)$\times 10^{-18}$} & {\Th{230}} & {75.4 ky} & {(1 -- 10)$\times 10^{-8}$} & 241.995 & (3 -- 30)$\times 10^{-10}$ \\
 & & & {\U{234}} & {245.5 ky} & {(2 -- 20)$\times 10^{-8}$} & \textit{295.2228} & (7 -- 70)$\times 10^{-10}$ \\
 & & & & & & \textit{\textbf{351.9321}} & (1 -- 10)$\times 10^{-9}$ \\
\hline
\multirow{4}{*}{\Sb{126}} & \multirow{4}{*}{12.35 days} & \multirow{4}{*}{(.2 -- 20)$\times 10^{-14}$} & \multirow{4}{*}{\Sn{126}} & \multirow{4}{*}{230 ky} & \multirow{4}{*}{(.1 -- 10)$\times 10^{-7}$} & \textit{\textbf{414.7}} & (.1 -- 10)$\times 10^{-8}$ \\
 & & & & & & \textit{\textbf{666.5}} & (.2 -- 10)$\times 10^{-8}$ \\
 & & & & & & \textit{\textbf{695.0}} & (.2 -- 10)$\times 10^{-8}$ \\
 & & & & & & \textit{\textbf{720.7}} & (.8 -- 80)$\times 10^{-9}$ \\
\hline
\multirow{3}{*}{\Sn{126}} & \multirow{3}{*}{230 ky} & \multirow{3}{*}{(.1 -- 10)$\times 10^{-7}$} & \multirow{3}{*}{K--T, Self} & \multirow{3}{*}{} & \multirow{3}{*}{} & \textbf{64.281} & (.1 -- 10)$\times 10^{-9}$ \\
 & & & & & & \textbf{86.938} &  (.1 -- 10)$\times 10^{-9}$ \\
 & & & & & & \textbf{87.567} & (.6 -- 50)$\times 10^{-9}$ \\
\hline
\end{tabular}
\label{tab:100ky_gammas}
\end{adjustwidth}
\end{table*}

\section{Discussion: Detectability Prospects}

To assess the detectability of kilonova remnants, we compare our results to a number of existing detectors as well as three different proposed satellite missions: \emph{Lunar Occultation Explorer (LOX)}, the \emph{Compton Spectrometer and Imager (COSI)}, and the \emph{All-sky Medium Energy Gamma-ray Observatory (AMEGO)}. Each of these missions has different strengths and weaknesses in observing kilonovae and their remnants and we review each of them here.  The most recent proposal for COSI is a COSI-SMEX mission and its sensitivity should lie between the COSI-X and GRX proposals (A.~Zoglauer, private communication).  We take the latest sensitivity curves from the \emph{AMEGO} \citep{rando17} and \emph{LOX} (R.~Miller, private communcation) collaborations.  The \emph{LOX} satellite is focused on the 0.1-few MeV range and its predicted sensitivities in this range are nearly two orders of magnitude lower than AMEGO, but it has only roughly 10\% energy resolution.

\begin{figure*}[htp]
\begin{tabular}{cc}
  \includegraphics[width=0.48\textwidth]{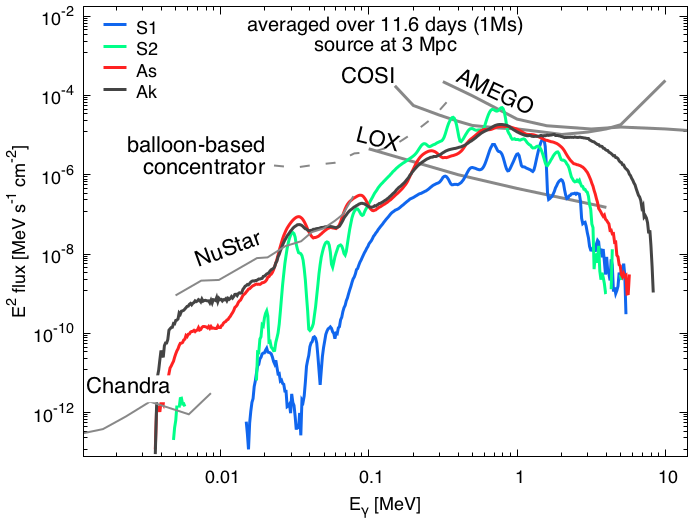} & 
  \includegraphics[width=0.48\textwidth]{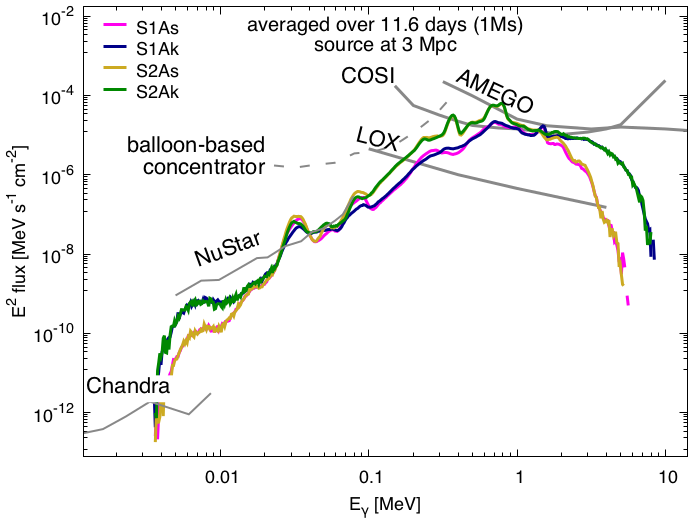}
  \\
  \includegraphics[width=0.48\textwidth]{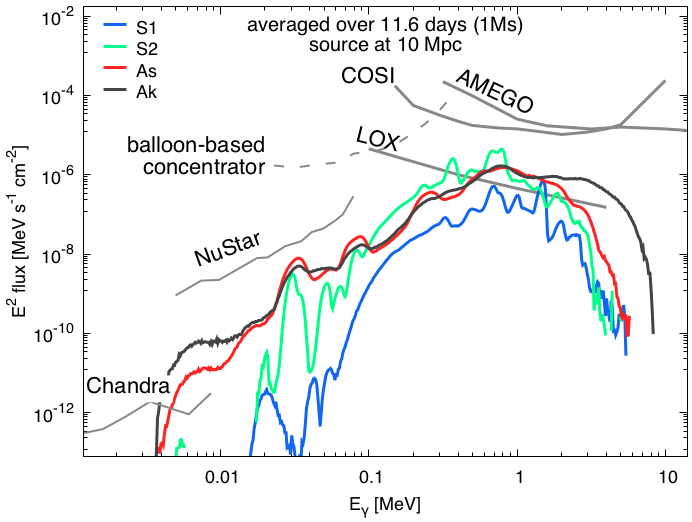} & 
  \includegraphics[width=0.48\textwidth]{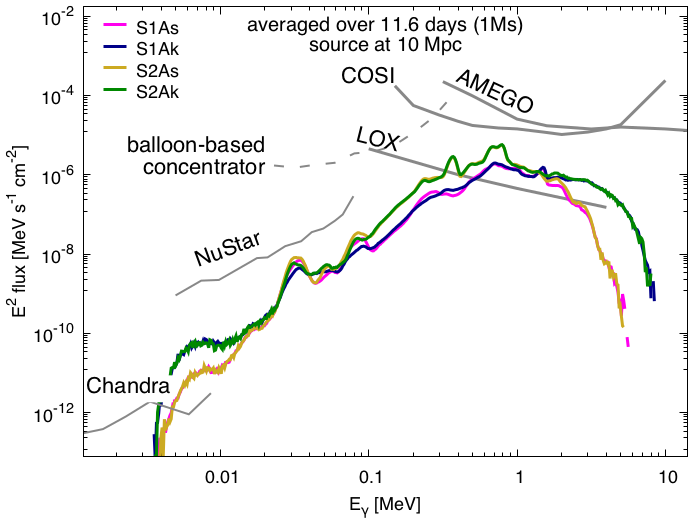}
\end{tabular}  
\caption{Synthetic spectra of one-component (left column) and two-component
   (right column) sources at distance 3~Mpc (top row) and 10~Mpc (bottom 
   row), integrated over the first 1Ms (11.6 days).
}
\label{fig:detectability1Ms}
\end{figure*}

Figure~\ref{fig:detectability1Ms} shows the transient signal, integrated over 1\,Ms starting from 1\,hour from the moment of merger for our ejecta models at 3~Mpc.  There are roughly 100 galaxies within 3\,Mpc and it is likely that the transient will be localized quickly, allowing a nearly instantaneous observation. Integrating over this period provides a reasonable estimate of the observed flux for these transients.  At the 10-100\,keV range, NuSTAR might be able to detect the signal from some mergers with 3\,Mpc.  For COSI-SMEX and AMEGO, telescopes that can be pointed, we assume a steady 1\,Ms observation.  Even assuming continuous observation by COSI-SMEX or AMEGO, such an event will be difficult to detect.  The LOX satellite should be able to detect a merger at 3\,Mpc, but a merger at 10\,Mpc will be just at the observing threshold.  We have also included the sensitivity of the balloon-based concentrator concept~\citep{shirazi18}.

\begin{figure*}[htp]
\begin{tabular}{cc}
 \includegraphics[width=0.48\textwidth]{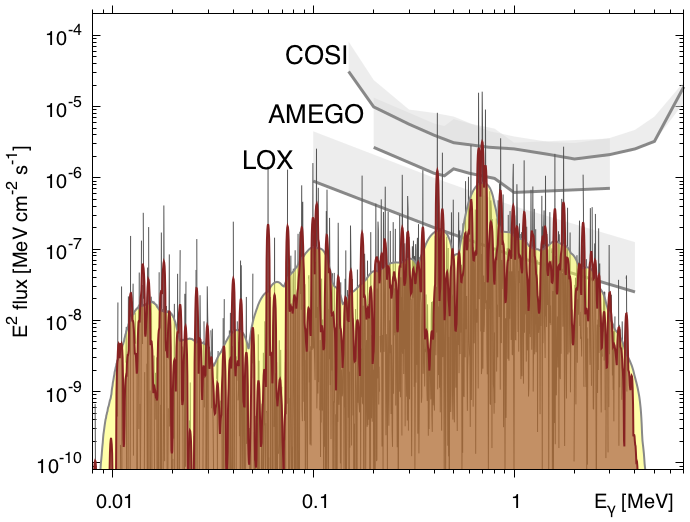} &
 \includegraphics[width=0.48\textwidth]{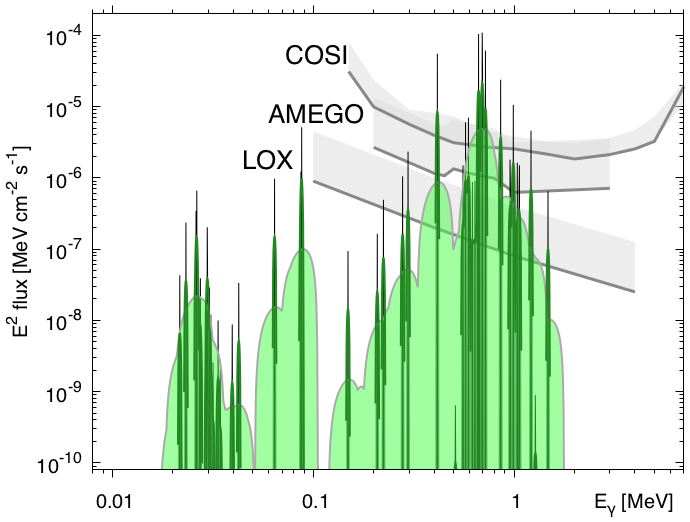}
\end{tabular}
\caption{Remnant $\gamma$-spectra for the high (left) and moderate
(right) initial neutron richness composition at ${10\;{\rm kyr}}$ epoch, 
compared with the \emph{LOX} sensitivity for one-year exposure (gray band).
Black thin lines represent simulated spectra, broadened with expansion 
velocities ${500\;{\rm km}\;{\rm s}^{-1}}$.
Dark- and light-shaded spectra correspond to further broadening to 
$1\%$ and $10\%$, emulating spectral sensitivity of the detector. 
Distance to the source is ${3\;{\rm kpc}}$.}
\label{fig:detectability1yrremn}
\end{figure*}

Probably more exciting is the possibility of a nearby, old kilonova remnant.  Figure~\ref{fig:detectability1yrremn} shows the detectability of a 10\,kyr remnant 3\,kpc from the Earth with sensitivities assuming 1 year of directed time.  With only a handful of merger remnants younger than 100\,kyr in the entire Milky Way, the odds of remnant this close to the Earth is less than 1\%.  A nearby remnant will likely be in a denser interstellar medium (e.g. $1 \, {\rm cm}^{-3}$), slowly expanding ($\sim 500 {\rm km}\,{\rm s}^{-1}$) and with a small extent ($1\,{\rm pc} \approx 0.5^\circ-1^\circ$).  Since these remnants vary slowly with time, sky surveys can be mined to look for this data (to achieve 1\,yr of directed time will require multiple years of telescope runtime).  With the high-energy resolution of COSI-SMEX it would be possible to distinguish the individual features in this signal, but would require a remnant less distant than 3\,kpc.  If we do not have to correct for the fact that nearby remnants can not be treated as point sources, \emph{LOX} will be able to observe these objects up to 10\,kpc and identify some of the largest features enough to distinguish between our two fission rate results.  Sensitivity and energy resolutions of \emph{AMEGO} \citep{kierans19} lies in between COSI-SMEX and \emph{LOX}.  

\section{Summary}

We have studied the \grays{} that arise primarily from the $\beta^-$ and $\alpha$-decays of radioactive isotopes in the kilonova ejecta of neutron star mergers\footnote{More work can be done to better understand the \gray{} signal and we have not included all sources of gamma-rays (e.g. fission-induced \gray{} emission).}. Comprehensive nucleosynthesis network modeling was used to generate the yields. We compared the signatures of a few representative compositions reflecting different types of neutron star merger ejecta and variations in nuclear physics. We studied both the transient kilonova phase that requires a new merger event, and an old neutron star merger remnant phase which could be identified in in our galaxy via its gamma emission. 

We further analyzed detectability prospects with upcoming telescope proposals for several existing (\emph{NuSTAR, Chandra}) and future (COSI, \emph{AMEGO}, \emph{LOX}) \gray{} missions. For the kilonova epoch, a neutron star merger event must happen within 10\,Mpc to be detectable (see Figure~\ref{fig:detectability1Ms}). If detected, it may be possible to distinguish broadened lines from very few individual isotopes (see Figure~\ref{fig:transient} and Table~\ref{tab:transient}) and reason about the ejecta composition. A nearby old merger remnant presents another potential detection possibility (see Figure~\ref{fig:detectability1yrremn}). For the remnant to be detectable with the high-energy resolution of COSI-SMEX it must be located closer than 3\,kpc. \emph{LOX} and \emph{AMEGO} will be able to observe such remnants up to 10\,kpc and identify some of the emitting isotopes. We demonstrated that in such case it is possible to discriminate between our two fission prescriptions (see Tables~\ref{tab:gammas} and \ref{tab:100ky_gammas}). Thus, if the spectrum can be measured, short-lived $\gamma$-emitters from fragments of actinide fission will allow to reason about fission models as well as potentially infer the amount of parent actinides.

We stress that a number of physics effects could alter the signals presented in this work. For example, synchrotron radiation may generate a background in the same energy range as our nuclear decay lines and residual thermal positron annihilation can produce a strong 511~keV feature~\citep{fuller19}. Furthermore, we have not included in our model the \grays{} from nuclear fission and nuclear isomeric states that are populated in radioactive decays, which may influence the observed spectrum. An example of an isomer that may have observable consequences for kilonovae is the \Nb{97} metastable state at 743~keV which has a 97.9\% $\gamma$ branch to the ground state. The de-excitation of this isomer, which is populated by the \bmd{} of \Zr{97}, may produce an observable feature near this energy beginning around 12 hours post-merger.

Just as with supernovae, \gray{} observations from the decay of radioactive nuclei require nearby events with a rate much lower than those achieved with optical and infrared observations.  But, as with supernovae, this study, along with the work of \cite{hotokezaka16} and \cite{wu19}, shows the unique potential of \grays{} to probe the details of nucleosynthesis in neutron star mergers (including nuclear physics), thereby ensuring their importance in understanding these powerful transients.

\acknowledgments
\section*{Acknowledgments}

We thank Andreas Zoglauer for providing data on upcoming gamma-ray detectors. O.K. is grateful to Brian Metzger for providing the original motivation for this work. We thank Almudena Arcones, Ryan Wollaeger, Kenta Hotokezaka, Masaomi Tanaka, Xilu Wang, Marius Eichler, Jonah Miller, Stephan Rosswog, Nicole Lloyd-Ronning, and Kei Davis for valuable discussions. We also thank the anonymous referee for constructive suggestions. This work was supported by the US Department of Energy through the Los Alamos National Laboratory. Los Alamos National Laboratory is operated by Triad National Security, LLC, for the National Nuclear Security Administration of U.S. Department of Energy (Contract No. 89233218CNA000001). Research presented in this article was supported by the Laboratory Directed Research and Development program of Los Alamos National Laboratory under project number 20190021DR. The work on the concentrator detector was supported by NASA grant NNX17AC85G.  All LANL calculations were performed on LANL Institutional Computing resources.  This work has benefited from support by the National Science Foundation under grant No. PHY-1430152 (JINA Center for the Evolution of the Elements).

\facilities{} 
\software{Maverick~\citep{hungerford03,hungerford05}, Numpy~\citep{vanderwalt11}, PRISM~\citep{mumpower17,mumpower18}}.


\begin{thebibliography}{}
\providecommand\natexlab[1]{#1}
\providecommand\JournalTitle[1]{#1}

\bibitem[{{Barnes} \& {Kasen}(2013)}]{barnes13}
{Barnes}, J., \& {Kasen}, D. 2013,
  \href{http://dx.doi.org/10.1088/0004-637X/775/1/18}{\JournalTitle{\apj}, 775,
  18}

\bibitem[{{Bauswein} {et~al.}(2013){Bauswein}, {Goriely}, \&
  {Janka}}]{bauswein13}
{Bauswein}, A., {Goriely}, S., \& {Janka}, H.-T. 2013,
  \href{http://dx.doi.org/10.1088/0004-637X/773/1/78}{\JournalTitle{\apj}, 773,
  78}

\bibitem[{Beers \& Christlieb(2005)}]{beers05}
Beers, T.~C., \& Christlieb, N. 2005,
  \href{http://dx.doi.org/10.1146/annurev.astro.42.053102.134057}{\JournalTitle{Annual
  Review of Astronomy and Astrophysics}, 43, 531}

\bibitem[{{Bloom} {et~al.}(1999){Bloom}, {Sigurdsson}, \& {Pols}}]{bloom99}
{Bloom}, J.~S., {Sigurdsson}, S., \& {Pols}, O.~R. 1999,
  \href{http://dx.doi.org/10.1046/j.1365-8711.1999.02437.x}{\JournalTitle{\mnras},
  305, 763}

\bibitem[{Brown {et~al.}(2018)Brown, Chadwick, Capote, Kahler, Trkov, Herman,
  Sonzogni, Danon, Carlson, Dunn, Smith, Hale, Arbanas, Arcilla, Bates, Beck,
  Becker, Brown, Casperson, Conlin, Cullen, Descalle, Firestone, Gaines, Guber,
  Hawari, Holmes, Johnson, Kawano, Kiedrowski, Koning, Kopecky, Leal, Lestone,
  Lubitz, Damián, Mattoon, McCutchan, Mughabghab, Navratil, Neudecker, Nobre,
  Noguere, Paris, Pigni, Plompen, Pritychenko, Pronyaev, Roubtsov, Rochman,
  Romano, Schillebeeckx, Simakov, Sin, Sirakov, Sleaford, Sobes, Soukhovitskii,
  Stetcu, Talou, Thompson, van~der Marck, Welser-Sherrill, Wiarda, White,
  Wormald, Wright, Zerkle, Žerovnik, \& Zhu}]{brown18}
Brown, D., Chadwick, M., Capote, R., {et~al.} 2018,
  \href{http://dx.doi.org/https://doi.org/10.1016/j.nds.2018.02.001}{\JournalTitle{Nuclear
  Data Sheets}, 148, 1 }, special Issue on Nuclear Reaction Data

\bibitem[{{Chen} \& {Holz}(2013)}]{chen13}
{Chen}, H.-Y., \& {Holz}, D.~E. 2013,
  \href{http://dx.doi.org/10.1103/PhysRevLett.111.181101}{\JournalTitle{Physical
  Review Letters}, 111, 181101}

\bibitem[{{Churazov} {et~al.}(2014){Churazov}, {Sunyaev}, {Isern},
  {Kn{\"o}dlseder}, {Jean}, {Lebrun}, {Chugai}, {Grebenev}, {Bravo}, {Sazonov},
  \& {Renaud}}]{churazov14}
{Churazov}, E., {Sunyaev}, R., {Isern}, J., {et~al.} 2014,
  \href{http://dx.doi.org/10.1038/nature13672}{\JournalTitle{\nat}, 512, 406}

\bibitem[{{C{\^o}t{\'e}} {et~al.}(2017){C{\^o}t{\'e}}, {Belczynski}, {Fryer},
  {Ritter}, {Paul}, {Wehmeyer}, \& {O'Shea}}]{cote17}
{C{\^o}t{\'e}}, B., {Belczynski}, K., {Fryer}, C.~L., {et~al.} 2017,
  \href{http://dx.doi.org/10.3847/1538-4357/aa5c8d}{\JournalTitle{\apj}, 836,
  230}

\bibitem[{{C{\^o}t{\'e}} {et~al.}(2018){C{\^o}t{\'e}}, {Fryer}, {Belczynski},
  {Korobkin}, {Chru{\'s}li{\'n}ska}, {Vassh}, {Mumpower}, {Lippuner},
  {Sprouse}, {Surman}, \& {Wollaeger}}]{cote18}
{C{\^o}t{\'e}}, B., {Fryer}, C.~L., {Belczynski}, K., {et~al.} 2018,
  \href{http://dx.doi.org/10.3847/1538-4357/aaad67}{\JournalTitle{\apj}, 855,
  99}

\bibitem[{{C{\^o}t{\'e}} {et~al.}(2019){C{\^o}t{\'e}}, {Eichler}, {Arcones},
  {Hansen}, {Simonetti}, {Frebel}, {Fryer}, {Pignatari}, {Reichert},
  {Belczynski}, \& {Matteucci}}]{cote19}
{C{\^o}t{\'e}}, B., {Eichler}, M., {Arcones}, A., {et~al.} 2019,
  \href{http://dx.doi.org/10.3847/1538-4357/ab10db}{\JournalTitle{\apj}, 875,
  106}

\bibitem[{{Dessart} {et~al.}(2009){Dessart}, {Ott}, {Burrows}, {Rosswog}, \&
  {Livne}}]{dessart09a}
{Dessart}, L., {Ott}, C.~D., {Burrows}, A., {Rosswog}, S., \& {Livne}, E. 2009,
  \href{http://dx.doi.org/10.1088/0004-637X/690/2/1681}{\JournalTitle{\apj},
  690, 1681}

\bibitem[{{Dominik} {et~al.}(2012){Dominik}, {Belczynski}, {Fryer}, {Holz},
  {Berti}, {Bulik}, {Mandel}, \& {O'Shaughnessy}}]{dominik12}
{Dominik}, M., {Belczynski}, K., {Fryer}, C., {et~al.} 2012,
  \href{http://dx.doi.org/10.1088/0004-637X/759/1/52}{\JournalTitle{\apj}, 759,
  52}

\bibitem[{{Eichler} {et~al.}(2019){Eichler}, {Sayar}, {Arcones}, \&
  {Rauscher}}]{eichler19}
{Eichler}, M., {Sayar}, W., {Arcones}, A., \& {Rauscher}, T. 2019,
  \href{http://dx.doi.org/10.3847/1538-4357/ab24cf}{\JournalTitle{\apj}, 879,
  47}

\bibitem[{{Endrizzi} {et~al.}(2016){Endrizzi}, {Ciolfi}, {Giacomazzo},
  {Kastaun}, \& {Kawamura}}]{endrizzi16}
{Endrizzi}, A., {Ciolfi}, R., {Giacomazzo}, B., {Kastaun}, W., \& {Kawamura},
  T. 2016,
  \href{http://dx.doi.org/10.1088/0264-9381/33/16/164001}{\JournalTitle{Classical
  and Quantum Gravity}, 33, 164001}

\bibitem[{{Evans} {et~al.}(2017){Evans}, {Cenko}, {Kennea}, {Emery}, {Kuin},
  {Korobkin}, {Wollaeger}, {Fryer}, {Madsen}, {Harrison}, {Xu}, {Nakar},
  {Hotokezaka}, {Lien}, {Campana}, {Oates}, {Troja}, {Breeveld}, {Marshall},
  {Barthelmy}, {Beardmore}, {Burrows}, {Cusumano}, {D'A{\`\i}}, {D'Avanzo},
  {D'Elia}, {de Pasquale}, {Even}, {Fontes}, {Forster}, {Garcia}, {Giommi},
  {Grefenstette}, {Gronwall}, {Hartmann}, {Heida}, {Hungerford}, {Kasliwal},
  {Krimm}, {Levan}, {Malesani}, {Melandri}, {Miyasaka}, {Nousek}, {O'Brien},
  {Osborne}, {Pagani}, {Page}, {Palmer}, {Perri}, {Pike}, {Racusin}, {Rosswog},
  {Siegel}, {Sakamoto}, {Sbarufatti}, {Tagliaferri}, {Tanvir}, \&
  {Tohuvavohu}}]{evans17}
{Evans}, P.~A., {Cenko}, S.~B., {Kennea}, J.~A., {et~al.} 2017,
  \href{http://dx.doi.org/10.1126/science.aap9580}{\JournalTitle{Science}, 358,
  1565}

\bibitem[{{Fern{\'a}ndez} {et~al.}(2017){Fern{\'a}ndez}, {Foucart}, {Kasen},
  {Lippuner}, {Desai}, \& {Roberts}}]{fernandez17}
{Fern{\'a}ndez}, R., {Foucart}, F., {Kasen}, D., {et~al.} 2017,
  \href{http://dx.doi.org/10.1088/1361-6382/aa7a77}{\JournalTitle{Classical and
  Quantum Gravity}, 34, 154001}

\bibitem[{{Fern{\'a}ndez} {et~al.}(2015){Fern{\'a}ndez}, {Kasen}, {Metzger}, \&
  {Quataert}}]{fernandez15}
{Fern{\'a}ndez}, R., {Kasen}, D., {Metzger}, B.~D., \& {Quataert}, E. 2015,
  \href{http://dx.doi.org/10.1093/mnras/stu2112}{\JournalTitle{\mnras}, 446,
  750}

\bibitem[{{Fern{\'a}ndez} \& {Metzger}(2013)}]{fernandez13}
{Fern{\'a}ndez}, R., \& {Metzger}, B.~D. 2013,
  \href{http://dx.doi.org/10.1093/mnras/stt1312}{\JournalTitle{\mnras}, 435,
  502}

\bibitem[{{Fong} \& {Berger}(2013)}]{fong13}
{Fong}, W., \& {Berger}, E. 2013,
  \href{http://dx.doi.org/10.1088/0004-637X/776/1/18}{\JournalTitle{\apj}, 776,
  18}

\bibitem[{{Fong} {et~al.}(2014){Fong}, {Berger}, {Metzger}, {Margutti},
  {Chornock}, {Migliori}, {Foley}, {Zauderer}, {Lunnan}, {Laskar}, {Desch},
  {Meech}, {Sonnett}, {Dickey}, {Hedlund}, \& {Harding}}]{fong14}
{Fong}, W., {Berger}, E., {Metzger}, B.~D., {et~al.} 2014,
  \href{http://dx.doi.org/10.1088/0004-637X/780/2/118}{\JournalTitle{\apj},
  780, 118}

\bibitem[{{Fontes} {et~al.}(2015){Fontes}, {Fryer}, {Hungerford}, {Hakel},
  {Colgan}, {Kilcrease}, \& {Sherrill}}]{fontes15}
{Fontes}, C.~J., {Fryer}, C.~L., {Hungerford}, A.~L., {et~al.} 2015,
  \href{http://dx.doi.org/10.1016/j.hedp.2015.06.002}{\JournalTitle{High Energy
  Density Physics}, 16, 53}

\bibitem[{{Fontes} {et~al.}(2019){Fontes}, {Fryer}, {Hungerford}, {Wollaeger},
  \& {Korobkin}}]{fontes19}
{Fontes}, C.~J., {Fryer}, C.~L., {Hungerford}, A.~L., {Wollaeger}, R.~T., \&
  {Korobkin}, O. 2019, \JournalTitle{arXiv e-prints}, arXiv:1904.08781

\bibitem[{{Fontes} {et~al.}(2017){Fontes}, {Fryer}, {Hungerford}, {Wollaeger},
  {Rosswog}, \& {Berger}}]{fontes17}
{Fontes}, C.~J., {Fryer}, C.~L., {Hungerford}, A.~L., {et~al.} 2017,
  \JournalTitle{arXiv e-prints},
  \href{http://arxiv.org/abs/1702.02990}{{\sffamily arXiv:1702.02990
  [astro-ph.HE]}}

\bibitem[{{Fryer} {et~al.}(1999{\natexlab{a}}){Fryer}, {Benz}, {Herant}, \&
  {Colgate}}]{fryer99c}
{Fryer}, C., {Benz}, W., {Herant}, M., \& {Colgate}, S.~A. 1999{\natexlab{a}},
  \href{http://dx.doi.org/10.1086/307119}{\JournalTitle{\apj}, 516, 892}

\bibitem[{{Fryer} {et~al.}(1999{\natexlab{b}}){Fryer}, {Woosley}, \&
  {Hartmann}}]{fryer99a}
{Fryer}, C.~L., {Woosley}, S.~E., \& {Hartmann}, D.~H. 1999{\natexlab{b}},
  \href{http://dx.doi.org/10.1086/307992}{\JournalTitle{\apj}, 526, 152}

\bibitem[{{Fuller} {et~al.}(2019){Fuller}, {Kusenko}, {Radice}, \&
  {Takhistov}}]{fuller19}
{Fuller}, G.~M., {Kusenko}, A., {Radice}, D., \& {Takhistov}, V. 2019,
  \href{http://dx.doi.org/10.1103/PhysRevLett.122.121101}{\JournalTitle{\prl},
  122, 121101}

\bibitem[{{Grefenstette} {et~al.}(2014){Grefenstette}, {Harrison}, {Boggs},
  {Reynolds}, {Fryer}, {Madsen}, {Wik}, {Zoglauer}, {Ellinger}, {Alexander},
  {An}, {Barret}, {Christensen}, {Craig}, {Forster}, {Giommi}, {Hailey},
  {Hornstrup}, {Kaspi}, {Kitaguchi}, {Koglin}, {Mao}, {Miyasaka}, {Mori},
  {Perri}, {Pivovaroff}, {Puccetti}, {Rana}, {Stern}, {Westergaard}, \&
  {Zhang}}]{grefenstette14}
{Grefenstette}, B.~W., {Harrison}, F.~A., {Boggs}, S.~E., {et~al.} 2014,
  \href{http://dx.doi.org/10.1038/nature12997}{\JournalTitle{\nat}, 506, 339}

\bibitem[{{Grefenstette} {et~al.}(2017){Grefenstette}, {Fryer}, {Harrison},
  {Boggs}, {DeLaney}, {Laming}, {Reynolds}, {Alexander}, {Barret},
  {Christensen}, {Craig}, {Forster}, {Giommi}, {Hailey}, {Hornstrup},
  {Kitaguchi}, {Koglin}, {Lopez}, {Mao}, {Madsen}, {Miyasaka}, {Mori}, {Perri},
  {Pivovaroff}, {Puccetti}, {Rana}, {Stern}, {Westergaard}, {Wik}, {Zhang}, \&
  {Zoglauer}}]{grefenstette17}
{Grefenstette}, B.~W., {Fryer}, C.~L., {Harrison}, F.~A., {et~al.} 2017,
  \href{http://dx.doi.org/10.3847/1538-4357/834/1/19}{\JournalTitle{\apj}, 834,
  19}

\bibitem[{{Grossman} {et~al.}(2014){Grossman}, {Korobkin}, {Rosswog}, \&
  {Piran}}]{grossman14}
{Grossman}, D., {Korobkin}, O., {Rosswog}, S., \& {Piran}, T. 2014,
  \href{http://dx.doi.org/10.1093/mnras/stt2503}{\JournalTitle{\mnras}, 439,
  757}

\bibitem[{Hansen {et~al.}(2018)Hansen, Holmbeck, Beers, Placco, Roederer,
  Frebel, Sakari, Simon, \& Thompson}]{hansen18}
Hansen, T.~T., Holmbeck, E.~M., Beers, T.~C., {et~al.} 2018,
  \href{http://dx.doi.org/10.3847/1538-4357/aabacc}{\JournalTitle{The
  Astrophysical Journal}, 858, 92}

\bibitem[{{Holmbeck} {et~al.}(2019){Holmbeck}, {Sprouse}, {Mumpower}, {Vassh},
  {Surman}, {Beers}, \& {Kawano}}]{holmbeck19}
{Holmbeck}, E.~M., {Sprouse}, T.~M., {Mumpower}, M.~R., {et~al.} 2019,
  \href{http://dx.doi.org/10.3847/1538-4357/aaefef}{\JournalTitle{\apj}, 870,
  23}

\bibitem[{{Horowitz} {et~al.}(2019){Horowitz}, {Arcones}, {C{\^o}t{\'e}},
  {Dillmann}, {Nazarewicz}, {Roederer}, {Schatz}, {Aprahamian}, {Atanasov},
  {Bauswein}, {Beers}, {Bliss}, {Brodeur}, {Clark}, {Frebel}, {Foucart},
  {Hansen}, {Just}, {Kankainen}, {McLaughlin}, {Kelly}, {Liddick}, {Lee},
  {Lippuner}, {Martin}, {Mendoza-Temis}, {Metzger}, {Mumpower}, {Perdikakis},
  {Pereira}, {O{\textquoteright}Shea}, {Reifarth}, {Rogers}, {Siegel},
  {Spyrou}, {Surman}, {Tang}, {Uesaka}, \& {Wang}}]{horowitz18}
{Horowitz}, C.~J., {Arcones}, A., {C{\^o}t{\'e}}, B., {et~al.} 2019,
  \href{http://dx.doi.org/10.1088/1361-6471/ab0849}{\JournalTitle{Journal of
  Physics G Nuclear Physics}, 46, 083001}

\bibitem[{{Hotokezaka} {et~al.}(2018){Hotokezaka}, {Beniamini}, \&
  {Piran}}]{hotokezaka18}
{Hotokezaka}, K., {Beniamini}, P., \& {Piran}, T. 2018,
  \href{http://dx.doi.org/10.1142/S0218271818420051}{\JournalTitle{International
  Journal of Modern Physics D}, 27, 1842005}

\bibitem[{{Hotokezaka} {et~al.}(2013){Hotokezaka}, {Kiuchi}, {Kyutoku},
  {Okawa}, {Sekiguchi}, {Shibata}, \& {Taniguchi}}]{hotokezaka13}
{Hotokezaka}, K., {Kiuchi}, K., {Kyutoku}, K., {et~al.} 2013,
  \href{http://dx.doi.org/10.1103/PhysRevD.87.024001}{\JournalTitle{\prd}, 87,
  024001}

\bibitem[{{Hotokezaka} {et~al.}(2017){Hotokezaka}, {Sari}, \&
  {Piran}}]{hotokezaka17}
{Hotokezaka}, K., {Sari}, R., \& {Piran}, T. 2017,
  \href{http://dx.doi.org/10.1093/mnras/stx411}{\JournalTitle{\mnras}, 468, 91}

\bibitem[{{Hotokezaka} {et~al.}(2016){Hotokezaka}, {Wanajo}, {Tanaka}, {Bamba},
  {Terada}, \& {Piran}}]{hotokezaka16}
{Hotokezaka}, K., {Wanajo}, S., {Tanaka}, M., {et~al.} 2016,
  \href{http://dx.doi.org/10.1093/mnras/stw404}{\JournalTitle{\mnras}, 459, 35}

\bibitem[{{Hungerford} {et~al.}(2005){Hungerford}, {Fryer}, \&
  {Rockefeller}}]{hungerford05}
{Hungerford}, A.~L., {Fryer}, C.~L., \& {Rockefeller}, G. 2005,
  \href{http://dx.doi.org/10.1086/497323}{\JournalTitle{\apj}, 635, 487}

\bibitem[{{Hungerford} {et~al.}(2003){Hungerford}, {Fryer}, \&
  {Warren}}]{hungerford03}
{Hungerford}, A.~L., {Fryer}, C.~L., \& {Warren}, M.~S. 2003,
  \href{http://dx.doi.org/10.1086/376776}{\JournalTitle{\apj}, 594, 390}

\bibitem[{{Janiuk}(2014)}]{janiuk14}
{Janiuk}, A. 2014,
  \href{http://dx.doi.org/10.1051/0004-6361/201423822}{\JournalTitle{\aap},
  568, A105}

\bibitem[{{Ji} {et~al.}(2019){Ji}, {Drout}, \& {Hansen}}]{ji19}
{Ji}, A.~P., {Drout}, M.~R., \& {Hansen}, T.~T. 2019,
  \href{http://dx.doi.org/10.3847/1538-4357/ab3291}{\JournalTitle{\apj}, 882,
  40}

\bibitem[{{Ji} \& {Frebel}(2018)}]{ji18}
{Ji}, A.~P., \& {Frebel}, A. 2018,
  \href{http://dx.doi.org/10.3847/1538-4357/aab14a}{\JournalTitle{\apj}, 856,
  138}

\bibitem[{{Jin} {et~al.}(2015){Jin}, {Li}, {Cano}, {Covino}, {Fan}, \&
  {Wei}}]{jin15}
{Jin}, Z.-P., {Li}, X., {Cano}, Z., {et~al.} 2015,
  \href{http://dx.doi.org/10.1088/2041-8205/811/2/L22}{\JournalTitle{\apjl},
  811, L22}

\bibitem[{{Jin} {et~al.}(2016){Jin}, {Hotokezaka}, {Li}, {Tanaka}, {D'Avanzo},
  {Fan}, {Covino}, {Wei}, \& {Piran}}]{jin16}
{Jin}, Z.-P., {Hotokezaka}, K., {Li}, X., {et~al.} 2016,
  \href{http://dx.doi.org/10.1038/ncomms12898}{\JournalTitle{Nature
  Communications}, 7, 12898}

\bibitem[{{Just} {et~al.}(2015){Just}, {Bauswein}, {Ardevol Pulpillo},
  {Goriely}, \& {Janka}}]{just15}
{Just}, O., {Bauswein}, A., {Ardevol Pulpillo}, R., {Goriely}, S., \& {Janka},
  H.-T. 2015,
  \href{http://dx.doi.org/10.1093/mnras/stv009}{\JournalTitle{\mnras}, 448,
  541}

\bibitem[{{Kalogera} {et~al.}(2004){Kalogera}, {Kim}, {Lorimer}, {Burgay},
  {D'Amico}, {Possenti}, {Manchester}, {Lyne}, {Joshi}, {McLaughlin}, {Kramer},
  {Sarkissian}, \& {Camilo}}]{kalogera04}
{Kalogera}, V., {Kim}, C., {Lorimer}, D.~R., {et~al.} 2004,
  \href{http://dx.doi.org/10.1086/382155}{\JournalTitle{\apjl}, 601, L179}

\bibitem[{{Kasen} {et~al.}(2013){Kasen}, {Badnell}, \& {Barnes}}]{kasen13}
{Kasen}, D., {Badnell}, N.~R., \& {Barnes}, J. 2013,
  \href{http://dx.doi.org/10.1088/0004-637X/774/1/25}{\JournalTitle{\apj}, 774,
  25}

\bibitem[{{Kasliwal} {et~al.}(2017){Kasliwal}, {Korobkin}, {Lau}, {Wollaeger},
  \& {Fryer}}]{kasliwal17}
{Kasliwal}, M.~M., {Korobkin}, O., {Lau}, R.~M., {Wollaeger}, R., \& {Fryer},
  C.~L. 2017,
  \href{http://dx.doi.org/10.3847/2041-8213/aa799d}{\JournalTitle{\apjl}, 843,
  L34}

\bibitem[{Kawano {et~al.}(2016)Kawano, Capote, Hilaire, \& Chau
  Huu-Tai}]{kawano16}
Kawano, T., Capote, R., Hilaire, S., \& Chau Huu-Tai, P. 2016,
  \href{http://dx.doi.org/10.1103/PhysRevC.94.014612}{\JournalTitle{Phys. Rev.
  C}, 94, 014612}

\bibitem[{{Kawano} {et~al.}(2008){Kawano}, {M{\"o}ller}, \&
  {Wilson}}]{kawano08}
{Kawano}, T., {M{\"o}ller}, P., \& {Wilson}, W.~B. 2008,
  \href{http://dx.doi.org/10.1103/PhysRevC.78.054601}{\JournalTitle{\prc}, 78,
  054601}

\bibitem[{Kierans(2019)}]{kierans19}
Kierans, C. 2019,
  \href{https://www.astro.unige.ch/integral2019/presentations}{in 12th INTEGRAL
  Conference and 1st AHEAD Gamma-Ray Workshop}

\bibitem[{{Kodama} \& {Takahashi}(1975)}]{kodama75}
{Kodama}, T., \& {Takahashi}, K. 1975,
  \href{http://dx.doi.org/10.1016/0375-9474(75)90381-4}{\JournalTitle{Nuclear
  Physics A}, 239, 489}

\bibitem[{{Korobkin} {et~al.}(2012){Korobkin}, {Rosswog}, {Arcones}, \&
  {Winteler}}]{korobkin12}
{Korobkin}, O., {Rosswog}, S., {Arcones}, A., \& {Winteler}, C. 2012,
  \href{http://dx.doi.org/10.1111/j.1365-2966.2012.21859.x}{\JournalTitle{\mnras},
  426, 1940}

\bibitem[{{Lamb} {et~al.}(2019){Lamb}, {Tanvir}, {Levan}, {de Ugarte Postigo},
  {Kawaguchi}, {Corsi}, {Evans}, {Gompertz}, {Malesani}, {Page}, {Wiersema},
  {Rosswog}, {Shibata}, {Tanaka}, {van der Horst}, {Cano}, {Fynbo}, {Fruchter},
  {Greiner}, {Heintz}, {Higgins}, {Hjorth}, {Izzo}, {Jakobsson}, {Kann},
  {O'Brien}, {Perley}, {Pian}, {Pugliese}, {Starling}, {Th{\"o}ne}, {Watson},
  {Wijers}, \& {Xu}}]{lamb19}
{Lamb}, G.~P., {Tanvir}, N.~R., {Levan}, A.~J., {et~al.} 2019,
  \href{http://dx.doi.org/10.3847/1538-4357/ab38bb}{\JournalTitle{\apj}, 883,
  48}

\bibitem[{{Lattimer} \& {Schramm}(1974)}]{lattimer74}
{Lattimer}, J.~M., \& {Schramm}, D.~N. 1974,
  \href{http://dx.doi.org/10.1086/181612}{\JournalTitle{\apjl}, 192, L145}

\bibitem[{{Li}(2019)}]{li19}
{Li}, L.-X. 2019,
  \href{http://dx.doi.org/10.3847/1538-4357/aaf961}{\JournalTitle{\apj}, 872,
  19}

\bibitem[{{Li} \& {Paczy{\'n}ski}(1998)}]{li98}
{Li}, L.-X., \& {Paczy{\'n}ski}, B. 1998,
  \href{http://dx.doi.org/10.1086/311680}{\JournalTitle{\apjl}, 507, L59}

\bibitem[{{Lippuner} \& {Roberts}(2015)}]{lippuner15}
{Lippuner}, J., \& {Roberts}, L.~F. 2015,
  \href{http://dx.doi.org/10.1088/0004-637X/815/2/82}{\JournalTitle{\apj}, 815,
  82}

\bibitem[{{Marley} {et~al.}(2013){Marley}, {Aprahamian}, {Mumpower}, {Nystrom},
  {Paul}, {Siegl}, {Strauss}, {Surman}, {Clark}, {Perez Galvan}, {Savard},
  {Morgan}, \& {Orford}}]{marley13}
{Marley}, S.~T., {Aprahamian}, A., {Mumpower}, M., {et~al.} 2013, in APS
  Meeting Abstracts, Vol. 2013, APS Division of Nuclear Physics Meeting
  Abstracts, JH.007

\bibitem[{{Martin} {et~al.}(2015){Martin}, {Perego}, {Arcones}, {Thielemann},
  {Korobkin}, \& {Rosswog}}]{martin15}
{Martin}, D., {Perego}, A., {Arcones}, A., {et~al.} 2015,
  \href{http://dx.doi.org/10.1088/0004-637X/813/1/2}{\JournalTitle{\apj}, 813,
  2}

\bibitem[{{Metzger} \& {Berger}(2012)}]{metzger12}
{Metzger}, B.~D., \& {Berger}, E. 2012,
  \href{http://dx.doi.org/10.1088/0004-637X/746/1/48}{\JournalTitle{\apj}, 746,
  48}

\bibitem[{{Metzger} {et~al.}(2010){Metzger}, {Mart{\'{\i}}nez-Pinedo},
  {Darbha}, {Quataert}, {Arcones}, {Kasen}, {Thomas}, {Nugent}, {Panov}, \&
  {Zinner}}]{metzger10a}
{Metzger}, B.~D., {Mart{\'{\i}}nez-Pinedo}, G., {Darbha}, S., {et~al.} 2010,
  \href{http://dx.doi.org/10.1111/j.1365-2966.2010.16864.x}{\JournalTitle{\mnras},
  406, 2650}

\bibitem[{{Miller} {et~al.}(2019){Miller}, {Ryan}, {Dolence}, {Burrows},
  {Fontes}, {Fryer}, {Korobkin}, {Lippuner}, {Mumpower}, \&
  {Wollaeger}}]{miller19}
{Miller}, J.~M., {Ryan}, B.~R., {Dolence}, J.~C., {et~al.} 2019,
  \href{http://dx.doi.org/10.1103/PhysRevD.100.023008}{\JournalTitle{\prd},
  100, 023008}

\bibitem[{{Milne} {et~al.}(2004){Milne}, {Hungerford}, {Fryer}, {Evans},
  {Urbatsch}, {Boggs}, {Isern}, {Bravo}, {Hirschmann}, {Kumagai}, {Pinto}, \&
  {The}}]{milne04}
{Milne}, P.~A., {Hungerford}, A.~L., {Fryer}, C.~L., {et~al.} 2004,
  \href{http://dx.doi.org/10.1086/423235}{\JournalTitle{\apj}, 613, 1101}

\bibitem[{{M{\"o}ller} {et~al.}(2019){M{\"o}ller}, {Mumpower}, {Kawano}, \&
  {Myers}}]{moller19}
{M{\"o}ller}, P., {Mumpower}, M.~R., {Kawano}, T., \& {Myers}, W.~D. 2019,
  \href{http://dx.doi.org/10.1016/j.adt.2018.03.003}{\JournalTitle{Atomic Data
  and Nuclear Data Tables}, 125, 1}

\bibitem[{{Mumpower} {et~al.}(2016){Mumpower}, {Kawano}, \&
  {M{\"o}ller}}]{mumpower16}
{Mumpower}, M.~R., {Kawano}, T., \& {M{\"o}ller}, P. 2016,
  \href{http://dx.doi.org/10.1103/PhysRevC.94.064317}{\JournalTitle{\prc}, 94,
  064317}

\bibitem[{{Mumpower} {et~al.}(2018){Mumpower}, {Kawano}, {Sprouse}, {Vassh},
  {Holmbeck}, {Surman}, \& {M{\"o}ller}}]{mumpower18}
{Mumpower}, M.~R., {Kawano}, T., {Sprouse}, T.~M., {et~al.} 2018,
  \href{http://dx.doi.org/10.3847/1538-4357/aaeaca}{\JournalTitle{\apj}, 869,
  14}

\bibitem[{{Mumpower} {et~al.}(2017){Mumpower}, {Kawano}, {Ullmann}, {Krti{\v
  c}ka}, \& {Sprouse}}]{mumpower17}
{Mumpower}, M.~R., {Kawano}, T., {Ullmann}, J.~L., {Krti{\v c}ka}, M., \&
  {Sprouse}, T.~M. 2017,
  \href{http://dx.doi.org/10.1103/PhysRevC.96.024612}{\JournalTitle{\prc}, 96,
  024612}

\bibitem[{{Mumpower} {et~al.}(2015){Mumpower}, {Surman}, {Fang}, {Beard},
  {M{\"o}ller}, {Kawano}, \& {Aprahamian}}]{mumpower15}
{Mumpower}, M.~R., {Surman}, R., {Fang}, D.-L., {et~al.} 2015,
  \href{http://dx.doi.org/10.1103/PhysRevC.92.035807}{\JournalTitle{\prc}, 92,
  035807}

\bibitem[{{Paul}(2018)}]{paul18}
{Paul}, D. 2018,
  \href{http://dx.doi.org/10.1093/mnras/sty840}{\JournalTitle{\mnras}, 477,
  4275}

\bibitem[{{Perego} {et~al.}(2014){Perego}, {Rosswog}, {Cabez{\'o}n},
  {Korobkin}, {K{\"a}ppeli}, {Arcones}, \& {Liebend{\"o}rfer}}]{perego14a}
{Perego}, A., {Rosswog}, S., {Cabez{\'o}n}, R.~M., {et~al.} 2014,
  \href{http://dx.doi.org/10.1093/mnras/stu1352}{\JournalTitle{\mnras}, 443,
  3134}

\bibitem[{{Perley} {et~al.}(2009){Perley}, {Metzger}, {Granot}, {Butler},
  {Sakamoto}, {Ramirez-Ruiz}, {Levan}, {Bloom}, {Miller}, {Bunker}, {Chen},
  {Filippenko}, {Gehrels}, {Glazebrook}, {Hall}, {Hurley}, {Kocevski}, {Li},
  {Lopez}, {Norris}, {Piro}, {Poznanski}, {Prochaska}, {Quataert}, \&
  {Tanvir}}]{perley09}
{Perley}, D.~A., {Metzger}, B.~D., {Granot}, J., {et~al.} 2009,
  \href{http://dx.doi.org/10.1088/0004-637X/696/2/1871}{\JournalTitle{\apj},
  696, 1871}

\bibitem[{{Pian} {et~al.}(2017){Pian}, {D'Avanzo}, {Benetti}, {Branchesi},
  {Brocato}, {Campana}, {Cappellaro}, {Covino}, {D'Elia}, {Fynbo}, {Getman},
  {Ghirlanda}, {Ghisellini}, {Grado}, {Greco}, {Hjorth}, {Kouveliotou},
  {Levan}, {Limatola}, {Malesani}, {Mazzali}, {Melandri}, {M{\o}ller},
  {Nicastro}, {Palazzi}, {Piranomonte}, {Rossi}, {Salafia}, {Selsing},
  {Stratta}, {Tanaka}, {Tanvir}, {Tomasella}, {Watson}, {Yang}, {Amati},
  {Antonelli}, {Ascenzi}, {Bernardini}, {Bo{\"e}r}, {Bufano}, {Bulgarelli},
  {Capaccioli}, {Casella}, {Castro-Tirado}, {Chassande-Mottin}, {Ciolfi},
  {Copperwheat}, {Dadina}, {De Cesare}, {di Paola}, {Fan}, {Gendre},
  {Giuffrida}, {Giunta}, {Hunt}, {Israel}, {Jin}, {Kasliwal}, {Klose}, {Lisi},
  {Longo}, {Maiorano}, {Mapelli}, {Masetti}, {Nava}, {Patricelli}, {Perley},
  {Pescalli}, {Piran}, {Possenti}, {Pulone}, {Razzano}, {Salvaterra},
  {Schipani}, {Spera}, {Stamerra}, {Stella}, {Tagliaferri}, {Testa}, {Troja},
  {Turatto}, {Vergani}, \& {Vergani}}]{pian17}
{Pian}, E., {D'Avanzo}, P., {Benetti}, S., {et~al.} 2017,
  \href{http://dx.doi.org/10.1038/nature24298}{\JournalTitle{\nat}, 551, 67}

\bibitem[{Piran(2005)}]{piran05}
Piran, T. 2005,
  \href{http://dx.doi.org/10.1103/RevModPhys.76.1143}{\JournalTitle{Rev. Mod.
  Phys.}, 76, 1143}

\bibitem[{{Piran} {et~al.}(2014){Piran}, {Korobkin}, \& {Rosswog}}]{piran14}
{Piran}, T., {Korobkin}, O., \& {Rosswog}, S. 2014, \JournalTitle{arXiv
  e-prints}, arXiv:1401.2166

\bibitem[{{Piran} {et~al.}(2013){Piran}, {Nakar}, \& {Rosswog}}]{piran13}
{Piran}, T., {Nakar}, E., \& {Rosswog}, S. 2013,
  \href{http://dx.doi.org/10.1093/mnras/stt037}{\JournalTitle{\mnras}, 430,
  2121}

\bibitem[{{Qian} {et~al.}(1998){Qian}, {Vogel}, \& {Wasserburg}}]{qian98}
{Qian}, Y.~Z., {Vogel}, P., \& {Wasserburg}, G.~J. 1998,
  \href{http://dx.doi.org/10.1086/306285}{\JournalTitle{\apj}, 506, 868}

\bibitem[{{Radice} {et~al.}(2016){Radice}, {Galeazzi}, {Lippuner}, {Roberts},
  {Ott}, \& {Rezzolla}}]{radice16}
{Radice}, D., {Galeazzi}, F., {Lippuner}, J., {et~al.} 2016,
  \href{http://dx.doi.org/10.1093/mnras/stw1227}{\JournalTitle{\mnras}, 460,
  3255}

\bibitem[{{Rando}(2017)}]{rando17}
{Rando}, R. 2017,
  \href{http://dx.doi.org/10.1088/1748-0221/12/11/C11024}{\JournalTitle{Journal
  of Instrumentation}, 12, C11024}

\bibitem[{{Ripley} {et~al.}(2014){Ripley}, {Metzger}, {Arcones}, \&
  {Mart{\'\i}nez-Pinedo}}]{ripley14}
{Ripley}, J.~L., {Metzger}, B.~D., {Arcones}, A., \& {Mart{\'\i}nez-Pinedo}, G.
  2014, \href{http://dx.doi.org/10.1093/mnras/stt2434}{\JournalTitle{\mnras},
  438, 3243}

\bibitem[{{Rosswog} {et~al.}(2014){Rosswog}, {Korobkin}, {Arcones},
  {Thielemann}, \& {Piran}}]{rosswog14}
{Rosswog}, S., {Korobkin}, O., {Arcones}, A., {Thielemann}, F.-K., \& {Piran},
  T. 2014,
  \href{http://dx.doi.org/10.1093/mnras/stt2502}{\JournalTitle{\mnras}, 439,
  744}

\bibitem[{{Rosswog} {et~al.}(2018){Rosswog}, {Sollerman}, {Feindt}, {Goobar},
  {Korobkin}, {Wollaeger}, {Fremling}, \& {Kasliwal}}]{rosswog18}
{Rosswog}, S., {Sollerman}, J., {Feindt}, U., {et~al.} 2018,
  \href{http://dx.doi.org/10.1051/0004-6361/201732117}{\JournalTitle{\aap},
  615, A132}

\bibitem[{{Sedov}(1946)}]{sedov46}
{Sedov}, L.~I. 1946, \JournalTitle{Appl. Math. Meth.}, 9, 294

\bibitem[{{Sekiguchi} {et~al.}(2016){Sekiguchi}, {Kiuchi}, {Kyutoku},
  {Shibata}, \& {Taniguchi}}]{sekiguchi16}
{Sekiguchi}, Y., {Kiuchi}, K., {Kyutoku}, K., {Shibata}, M., \& {Taniguchi}, K.
  2016,
  \href{http://dx.doi.org/10.1103/PhysRevD.93.124046}{\JournalTitle{\prd}, 93,
  124046}

\bibitem[{{Shirazi} {et~al.}(2018){Shirazi}, {Bloser}, {Krzanowskic}, {Legere},
  \& {McConnell}}]{shirazi18}
{Shirazi}, F., {Bloser}, P.~F., {Krzanowskic}, J.~E., {Legere}, J.~S., \&
  {McConnell}, M.~L. 2018, \href{http://dx.doi.org/10.1117/12.2312300}{in
  Society of Photo-Optical Instrumentation Engineers (SPIE) Conference Series,
  Vol. 10699, Space Telescopes and Instrumentation 2018: Ultraviolet to Gamma
  Ray}, 106995V

\bibitem[{{Siegel} \& {Metzger}(2017)}]{siegel17}
{Siegel}, D.~M., \& {Metzger}, B.~D. 2017,
  \href{http://dx.doi.org/10.1103/PhysRevLett.119.231102}{\JournalTitle{\prl},
  119, 231102}

\bibitem[{{Simonetti} {et~al.}(2019){Simonetti}, {Matteucci}, {Greggio}, \&
  {Cescutti}}]{simonetti19}
{Simonetti}, P., {Matteucci}, F., {Greggio}, L., \& {Cescutti}, G. 2019,
  \href{http://dx.doi.org/10.1093/mnras/stz991}{\JournalTitle{\mnras}, 486,
  2896}

\bibitem[{{Sneden} {et~al.}(1996){Sneden}, {McWilliam}, {Preston}, {Cowan},
  {Burris}, \& {Armosky}}]{sneden96}
{Sneden}, C., {McWilliam}, A., {Preston}, G.~W., {et~al.} 1996,
  \href{http://dx.doi.org/10.1086/177656}{\JournalTitle{\apj}, 467, 819}

\bibitem[{{Sprouse} {et~al.}(2019){Sprouse}, {Navarro Perez}, {Surman},
  {Mumpower}, {McLaughlin}, \& {Schunck}}]{sprouse19}
{Sprouse}, T.~M., {Navarro Perez}, R., {Surman}, R., {et~al.} 2019,
  \JournalTitle{arXiv e-prints},
  \href{http://arxiv.org/abs/1901.10337}{{\sffamily arXiv:1901.10337
  [nucl-th]}}

\bibitem[{Spyrou {et~al.}(2016)Spyrou, Liddick, Naqvi, Crider, Dombos, Bleuel,
  Brown, Couture, Crespo~Campo, Guttormsen, Larsen, Lewis, M\"oller, Mosby,
  Mumpower, Perdikakis, Prokop, Renstr\o{}m, Siem, Quinn, \&
  Valenta}]{spyrou16}
Spyrou, A., Liddick, S.~N., Naqvi, F., {et~al.} 2016,
  \href{http://dx.doi.org/10.1103/PhysRevLett.117.142701}{\JournalTitle{Phys.
  Rev. Lett.}, 117, 142701}

\bibitem[{{Tanaka} \& {Hotokezaka}(2013)}]{tanaka13}
{Tanaka}, M., \& {Hotokezaka}, K. 2013,
  \href{http://dx.doi.org/10.1088/0004-637X/775/2/113}{\JournalTitle{\apj},
  775, 113}

\bibitem[{{Tanvir} {et~al.}(2013){Tanvir}, {Levan}, {Fruchter}, {Hjorth},
  {Hounsell}, {Wiersema}, \& {Tunnicliffe}}]{tanvir13}
{Tanvir}, N.~R., {Levan}, A.~J., {Fruchter}, A.~S., {et~al.} 2013,
  \href{http://dx.doi.org/10.1038/nature12505}{\JournalTitle{\nat}, 500, 547}

\bibitem[{{Tanvir} {et~al.}(2017){Tanvir}, {Levan},
  {Gonz{\'a}lez-Fern{\'a}ndez}, {Korobkin}, {Mandel}, {Rosswog}, {Hjorth},
  {D'Avanzo}, {Fruchter}, {Fryer}, {Kangas}, {Milvang-Jensen}, {Rosetti},
  {Steeghs}, {Wollaeger}, {Cano}, {Copperwheat}, {Covino}, {D'Elia}, {de Ugarte
  Postigo}, {Evans}, {Even}, {Fairhurst}, {Figuera Jaimes}, {Fontes}, {Fujii},
  {Fynbo}, {Gompertz}, {Greiner}, {Hodosan}, {Irwin}, {Jakobsson},
  {J{\o}rgensen}, {Kann}, {Lyman}, {Malesani}, {McMahon}, {Melandri},
  {O'Brien}, {Osborne}, {Palazzi}, {Perley}, {Pian}, {Piranomonte}, {Rabus},
  {Rol}, {Rowlinson}, {Schulze}, {Sutton}, {Th{\"o}ne}, {Ulaczyk}, {Watson},
  {Wiersema}, \& {Wijers}}]{tanvir17}
{Tanvir}, N.~R., {Levan}, A.~J., {Gonz{\'a}lez-Fern{\'a}ndez}, C., {et~al.}
  2017, \href{http://dx.doi.org/10.3847/2041-8213/aa90b6}{\JournalTitle{\apjl},
  848, L27}

\bibitem[{{Taylor}(1950)}]{taylor50}
{Taylor}, G. 1950,
  \href{http://dx.doi.org/10.1098/rspa.1950.0049}{\JournalTitle{Proceedings of
  the Royal Society of London Series A}, 201, 159}

\bibitem[{Taylor(1941)}]{taylor41}
Taylor, G.~I. 1941, The formation of a blast wave by a very intense explosion,
  Tech. Rep. Brit Rept RC-210, National Defense Research Committee (NDRC), Div.
  B, reprinted in (1950) Proc Roy Soc A 186:159

\bibitem[{{The} \& {Burrows}(2014)}]{the14}
{The}, L.-S., \& {Burrows}, A. 2014,
  \href{http://dx.doi.org/10.1088/0004-637X/786/2/141}{\JournalTitle{\apj},
  786, 141}

\bibitem[{{Troja} {et~al.}(2017){Troja}, {Piro}, {van Eerten}, {Wollaeger},
  {Im}, {Fox}, {Butler}, {Cenko}, {Sakamoto}, {Fryer}, {Ricci}, {Lien}, {Ryan},
  {Korobkin}, {Lee}, {Burgess}, {Lee}, {Watson}, {Choi}, {Covino}, {D'Avanzo},
  {Fontes}, {Gonz{\'a}lez}, {Khandrika}, {Kim}, {Kim}, {Lee}, {Lee}, {Kutyrev},
  {Lim}, {S{\'a}nchez-Ram{\'{\i}}rez}, {Veilleux}, {Wieringa}, \&
  {Yoon}}]{troja17}
{Troja}, E., {Piro}, L., {van Eerten}, H., {et~al.} 2017,
  \href{http://dx.doi.org/10.1038/nature24290}{\JournalTitle{\nat}, 551, 71}

\bibitem[{{van der Walt} {et~al.}(2011){van der Walt}, {Colbert}, \&
  {Varoquaux}}]{vanderwalt11}
{van der Walt}, S., {Colbert}, S.~C., \& {Varoquaux}, G. 2011,
  \href{http://dx.doi.org/10.1109/MCSE.2011.37}{\JournalTitle{Computing in
  Science and Engineering}, 13, 22}

\bibitem[{{Van Schelt} {et~al.}(2013){Van Schelt}, {Lascar}, {Savard}, {Clark},
  {Bertone}, {Caldwell}, {Chaudhuri}, {Levand }, {Li}, {Morgan}, {Orford},
  {Segel}, {Sharma}, \& {Sternberg}}]{vanschelt13}
{Van Schelt}, J., {Lascar}, D., {Savard}, G., {et~al.} 2013,
  \href{http://dx.doi.org/10.1103/PhysRevLett.111.061102}{\JournalTitle{\prl},
  111, 061102}

\bibitem[{{Vassh} {et~al.}(2019){Vassh}, {Vogt}, {Surman}, {Randrup},
  {Sprouse}, {Mumpower}, {Jaffke}, {Shaw}, {Holmbeck}, {Zhu}, \&
  {McLaughlin}}]{vassh19}
{Vassh}, N., {Vogt}, R., {Surman}, R., {et~al.} 2019,
  \href{http://dx.doi.org/10.1088/1361-6471/ab0bea}{\JournalTitle{Journal of
  Physics G Nuclear Physics}, 46, 065202}

\bibitem[{{Wanajo} {et~al.}(2014){Wanajo}, {Sekiguchi}, {Nishimura}, {Kiuchi},
  {Kyutoku}, \& {Shibata}}]{wanajo14}
{Wanajo}, S., {Sekiguchi}, Y., {Nishimura}, N., {et~al.} 2014,
  \href{http://dx.doi.org/10.1088/2041-8205/789/2/L39}{\JournalTitle{\apjl},
  789, L39}

\bibitem[{{Watson} {et~al.}(2019){Watson}, {Hansen}, {Selsing}, {Koch},
  {Malesani}, {Andersen}, {Fynbo}, {Arcones}, {Bauswein}, {Covino}, {Grado},
  {Heintz}, {Hunt}, {Kouveliotou}, {Leloudas}, {Levan}, {Mazzali}, \&
  {Pian}}]{watson19}
{Watson}, D., {Hansen}, C.~J., {Selsing}, J., {et~al.} 2019,
  \href{http://dx.doi.org/10.1038/s41586-019-1676-3}{\JournalTitle{\nat}, 574,
  497}

\bibitem[{{Wiggins} {et~al.}(2018){Wiggins}, {Fryer}, {Smidt}, {Hartmann},
  {Lloyd-Ronning}, \& {Belcynski}}]{wiggins18}
{Wiggins}, B.~K., {Fryer}, C.~L., {Smidt}, J.~M., {et~al.} 2018,
  \href{http://dx.doi.org/10.3847/1538-4357/aad2d4}{\JournalTitle{\apj}, 865,
  27}

\bibitem[{{Wollaeger} {et~al.}(2018){Wollaeger}, {Korobkin}, {Fontes},
  {Rosswog}, {Even}, {Fryer}, {Sollerman}, {Hungerford}, {van Rossum}, \&
  {Wollaber}}]{wollaeger18}
{Wollaeger}, R.~T., {Korobkin}, O., {Fontes}, C.~J., {et~al.} 2018,
  \href{http://dx.doi.org/10.1093/mnras/sty1018}{\JournalTitle{\mnras}, 478,
  3298}

\bibitem[{{Wollaeger} {et~al.}(2019){Wollaeger}, {Fryer}, {Fontes}, {Lippuner},
  {Vestrand}, {Mumpower}, {Korobkin}, {Hungerford}, \& {Even}}]{wollaeger19}
{Wollaeger}, R.~T., {Fryer}, C.~L., {Fontes}, C.~J., {et~al.} 2019,
  \href{http://dx.doi.org/10.3847/1538-4357/ab25f5}{\JournalTitle{\apj}, 880,
  22}

\bibitem[{{Wu} {et~al.}(2019){Wu}, {Banerjee}, {Metzger},
  {Mart{\'\i}nez-Pinedo}, {Aramaki}, {Burns}, {Hailey}, {Barnes}, \&
  {Karagiorgi}}]{wu19}
{Wu}, M.-R., {Banerjee}, P., {Metzger}, B.~D., {et~al.} 2019,
  \href{http://dx.doi.org/10.3847/1538-4357/ab2593}{\JournalTitle{\apj}, 880,
  23}

\bibitem[{Yokoyama {et~al.}(2019)Yokoyama, Grzywacz, Rasco, Brewer,
  Rykaczewski, Dillmann, Tain, Nishimura, Ahn, Algora, Allmond, Agramunt, Baba,
  Bae, Bruno, Caballero-Folch, Calvino, Coleman-Smith, Cortes, Davinson,
  Domingo-Pardo, Estrade, Fukuda, Go, Griffin, Ha, Hall, Harkness-Brennan,
  Heideman, Isobe, Kahl, Karny, Kawano, Khiem, King, Kiss, Korgul, Kubono,
  Labiche, Lazarus, Liang, Liu, Lorusso, Madurga, Matsui, Miernik, Montes,
  Morales, Morrall, Nepal, Page, Phong, Piersa, Prydderch, Pucknell, Rajabali,
  Rubio, Saito, Sakurai, Shimizu, Simpson, Singh, Stracener, Sumikama, Surman,
  Suzuki, Takeda, Tarife\~no Saldivia, Thomas, Tolosa-Delgado, Woli\ifmmode
  \acute{n}\else~\'{n}\fi{}ska Cichocka, Woods, \& Xu}]{yokoyama19}
Yokoyama, R., Grzywacz, R., Rasco, B.~C., {et~al.} 2019,
  \href{http://dx.doi.org/10.1103/PhysRevC.100.031302}{\JournalTitle{Phys. Rev.
  C}, 100, 031302}

\end{thebibliography}
\end{document}